\documentclass[
 reprint,
notitlepage,
 amsmath,amssymb,
 aps,
 onecolumn,
]{revtex4-1}

\usepackage{graphicx}
\usepackage{dcolumn}
\usepackage{color}
\usepackage{subfigure}
\usepackage{amsmath}
\usepackage{amssymb} 
\usepackage{color}
\usepackage[normalem]{ulem}
\usepackage{bm}
\usepackage{natbib}
\begin{document}

\title{Erosion of cohesive grains by an impinging turbulent jet}

\author{Ram Sudhir Sharma,$^{1}$ Mingze Gong,$^{1}$  Sivar Azadi,$^{1}$ Adrien Gans,$^{2,3}$ Philippe Gondret,$^{4}$ and Alban Sauret$^{1,}$}\email{asauret@ucsb.edu}
\affiliation{$^{1}$ Department of Mechanical Engineering, University of California, Santa Barbara, CA 93106, USA \\
$^{2}$ Universit\'e Aix-Marseille, CNRS, IUSTI,  13453 Marseille, France \\
$^{3}$ Universit\'e Rennes, CNRS, IPR, UMR 6251, F-35000 Rennes, France \\
$^{4}$ Universit\'e Paris-Saclay, CNRS, Laboratoire FAST, 91405 Orsay, France}

\date{\today}

\begin{abstract}
The erosion and transport of particles by an impinging turbulent jet in air is observed in various situations, such as the cleaning of a surface or during the landing of a spacecraft. The presence of inter-particle cohesive forces modifies the erosion threshold, beyond which grains are transported. The cohesion also influences the resulting formation and shape of the crater. In this paper, we characterize the role of the cohesive forces on the erosion of a flat granular bed by an impinging normal turbulent jet in air. We perform experiments using a cohesion-controlled granular material to finely tune the cohesion between particles while keeping the other properties constant. We investigate the effects of the cohesion on the erosion threshold and show that the results can be rationalized by a cohesive Shields number that accounts for the inter-particles cohesion force. Despite the complex nature of a turbulent jet, we can provide a scaling law to correlate the jet erosion threshold, based on the outlet velocity at the nozzle, to a local cohesive Shields number. The presence of cohesion between the grains also modifies the shape of the resulting crater, the transport of grains, and the local erosion process.
\end{abstract}

\maketitle

\section{Introduction} \label{sec:intro}

The erosion of granular soils is ubiquitous in many fields ranging from environmental science \cite{partheniades1965erosion,lachaussee2018competitive} to aerospace engineering \cite{bousser2014solid}. In these applications, soil stability and the modification of the local topography when the soil is subjected to a fluid stress is a significant issue \cite{wan2004investigation}. For example, this situation is observed when a rocket takes off or lands due to the turbulent jet associated with the propulsion \cite{metzger2011phenomenology}. This situation is also encountered during measurements of soil cohesion prior to the construction of civil structures \cite{baets2008modelling}. Besides, erosion and removal of particles by a turbulent jet are of great interest in clean-up processes, such as at nuclear sites that use this process to rid reactor surfaces of harmful particles \cite{kohli2018developments}.

The erosion of non-cohesive grains by a fluid flow is governed by the balance of the local shear stress $\tau_{\rm f}$ exerted by the flow, which is the driving mechanism of erosion and grain transport, and the gravity force acting on a grain, which tends to prevent erosion through a friction force. The initiation of the grain motion is thus described by the ratio of the force exerted by the fluid, which scales as $\tau_{\rm f}\,d^2$, and the apparent weight of the grain, scaling as $(\rho_{\rm g}-\rho_{\rm f})\, g\, d^{3}$, where $\rho_{\rm g}$ and $\rho_{\rm f}$ are the density of the granular material and the fluid, respectively, and $d$ is the grain diameter. The ratio of these two forces leads to the Shields number: ${\rm Sh}=\tau_{\rm f}/[(\rho_{\rm g}-\rho_{\rm f})\, g\, d]$ \cite{shields1936anwendung,guo1997discussion,andreotti2013granular}. The threshold value of the Shields number, beyond which erosion is triggered, depends on the nature of the flow, laminar or turbulent, and on the particle Reynolds number, ${\rm Re}_{\rm p}=u_\ell\,d/\nu$, where $u_\ell$ is the characteristic local velocity of the flow, and $\nu$ is the kinematic viscosity of the fluid \cite{guo2002hunter,cao2006explicit}. Beyond the erosion threshold, the grains are transported by the flow through rolling, saltation, and suspension \cite{van1984sediment}. The flux of transported grains depends on the difference between the local Shields number and the threshold Shields number for the granular bed considered \cite{andreotti2013granular}.

A large part of the studies on erosion has considered the erosion of grains subjected to a unidirectional and homogeneous (translation invariant) tangential flow \cite{brach1988impact,williams2001erosion,guidoux2010contact}. This configuration is relevant for dune formation in the desert \cite{Andreotti2006PRL,Laity2009,Kok2012,charru2013sand,gunn2022sets}, or for sediment transport in rivers \cite{bungartz2004significance,Seminara2010,Phillips2019,Popovic2021PNAS}. Although this simple flow configuration allows obtaining relevant scaling laws for sediment erosion and transport, various configurations involve a granular bed subjected to a flow that is not homogeneous. In this case, the erosion can become localized, and it is possible to observe, for instance, the formation of a crater. A relevant configuration for the cleaning of surfaces consists in impacting a turbulent jet on a surface to remove the dust particles \cite{du2019turbulent}. A related configuration is also used in civil engineering during a jet erosion test, which consists of impacting a jet perpendicularly to a surface and measuring the depth eroded by the jet over time \cite{regazzoni2011investigation,karamigolbaghi2017critical}. Using empirical laws, it is then possible to obtain information on the erodibility of the sediment layer \cite{daly2013scour}. To refine the empirical models, different studies have considered laboratory configurations of a jet impinging normally on a granular bed \cite{badr2014erosion,badr2016crater,brunier2017erosion,haddadchi2018alternative,brunier2020generalized}. Various model studies on non-cohesive granular media in air or underwater have shown that the erosion threshold can be predicted using a free jet model, taking into account the position of the virtual origin of the jet \cite{badr2014erosion,brunier2017erosion}.  Numerical studies for laminar jets have also shown the relevance of the Shields number to describe the erosion threshold of the granular bed \cite{derksen2011simulations,benseghier2020relevance}. Beyond the erosion threshold, \textit{i.e.}, for larger Shields numbers, different morphologies of craters were highlighted, from shallow parabolic craters (type I) to deep conical craters (type II) \cite{badr2016crater,lamarche2015cratering,071.Guleria-2020}. These different morphologies depend on the distance of the jet from the granular bed, the velocity of the jet, and the size of the particles. Besides, the feedback of the crater shape on the turbulent flow plays an important role on the final morphology \cite{badr2016crater}.

Most model studies devoted to the erosion of a granular bed by a turbulent jet have considered cohesionless granular material, usually glass beads. As a result, a physical picture of the influence of the cohesion between the grains on the erosion threshold and on the shape of the asymptotic crater, \textit{i.e.}, the steady morphology observed at long time, remains more elusive. In particular, a turbulent jet brings an additional complexity due to the complex structure of the flow. Inter-particle cohesion is especially relevant for natural soils but remains challenging to control and quantify. Recently, an experimental study has considered the erosion by an immersed laminar jet of a cohesive granular bed using solid cohesive bonds between millimetric grains \cite{brunier2020generalized}. The use of capillary liquid bridges between the grains is not suitable for erosion experiments. Indeed, liquid bonds can be drained when subject to a flow so that the local cohesion of the material depends on the application time of the hydrodynamic stress. The drawback with solid bonds is that once the erosion threshold is reached, the bonds break irreversibly and the grains are not cohesive anymore \cite{brunier2020generalized}. Therefore, this approach does not reproduce configurations where the inter-particle force between particles can be restored after breakup. 

In this study, we address the situation of cohesive grains where the bonds can be formed again after the erosion occurs. To do so, we rely on an experimental method developed recently to create a cohesion-controlled granular material (CCGM) \cite{gans2020cohesion}. We consider the erosion of a flat granular bed made of cohesive spherical beads by an impinging turbulent jet in air. The influence of the cohesive force between the grains is captured by the cohesion number, corresponding to the ratio of the cohesive force $F_{\rm c}$ and the gravitational force acting on the particle $F_{\rm w}$: ${\rm Co}=F_{\rm c}/F_{\rm w}$. The CCGM allows us to tune finely the cohesive force $F_{\rm c}$, thus varying the cohesion number while keeping all other parameters, such as the grain size, constant. Such an approach allows us to extract the specific contribution of cohesion to the erosion process. We present in section \ref{sec:setup} the experimental methods and the CCGM used in this study. We then focus on the erosion threshold in section \ref{sec:erosion_threshold}, first considering cohesionless grains (${\rm Co}=0$) and then adding cohesion between particles (${\rm Co} > 0$). We show that the erosion threshold can be rescaled when accounting for the additional force induced by the cohesive forces. We further demonstrate that this model works both at the scale of the jet as well as the particle-scale. We then consider in section \ref{sec:crater_morphology} the change in the morphology of the asymptotic crater due to cohesion, as well as the dynamics of the formation of the crater.

\section{Experimental methods} \label{sec:setup}

\subsection{Experimental setup and cohesion-controlled granular material}

\begin{figure}
\centering
\subfigure[]{\includegraphics[width = 0.45\textwidth]{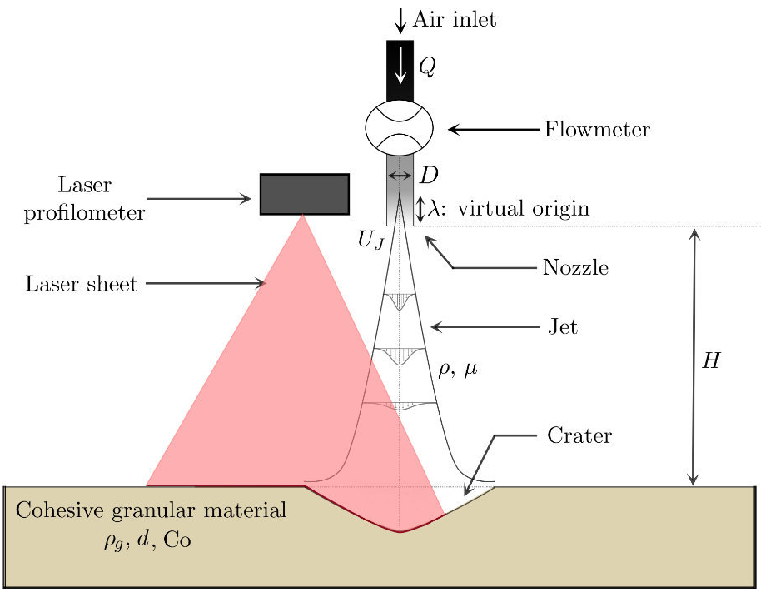}}
\subfigure[]{\includegraphics[width = 0.45\textwidth]{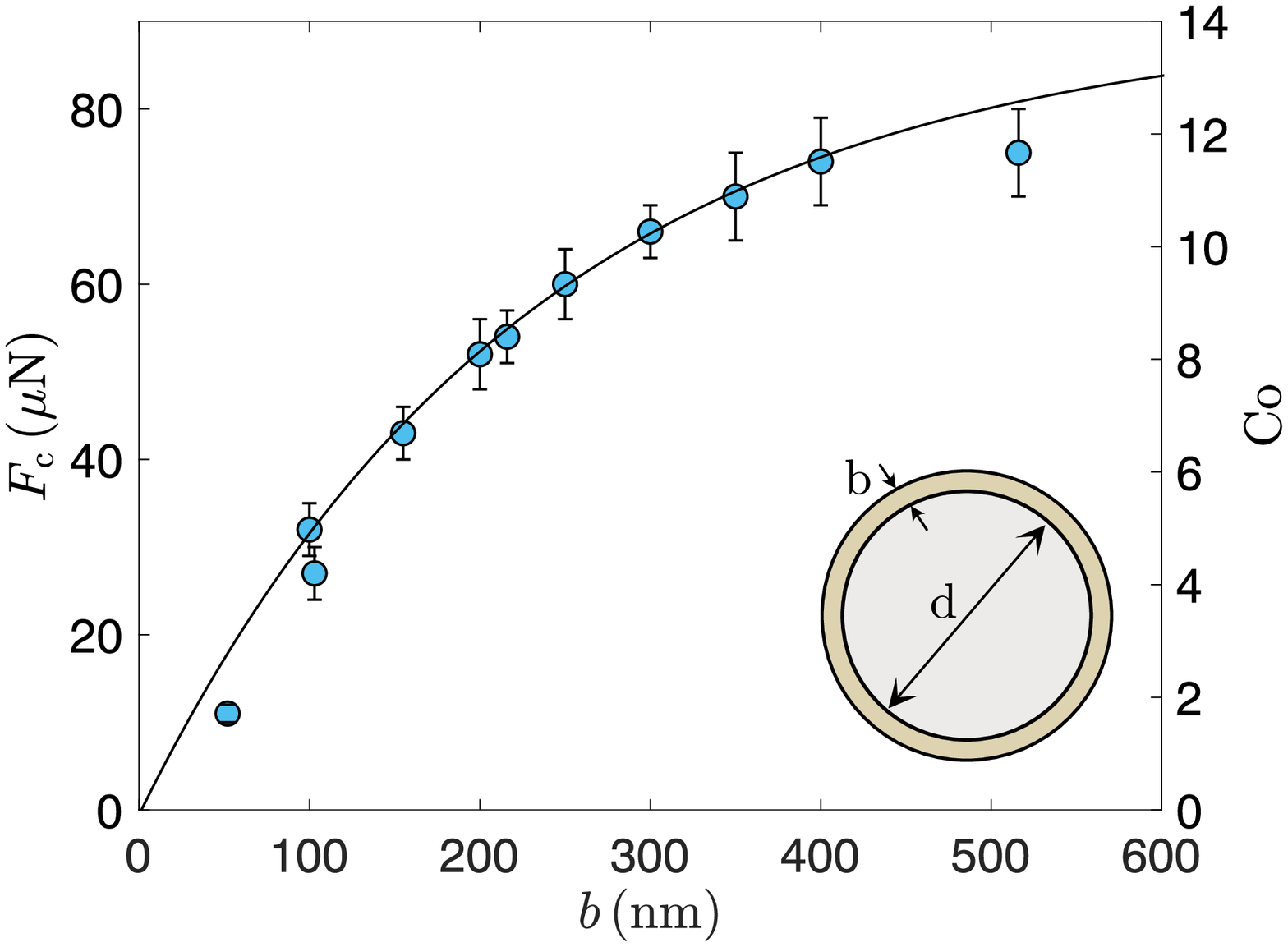}} \\
\subfigure[]{\includegraphics[width = 0.42\textwidth]{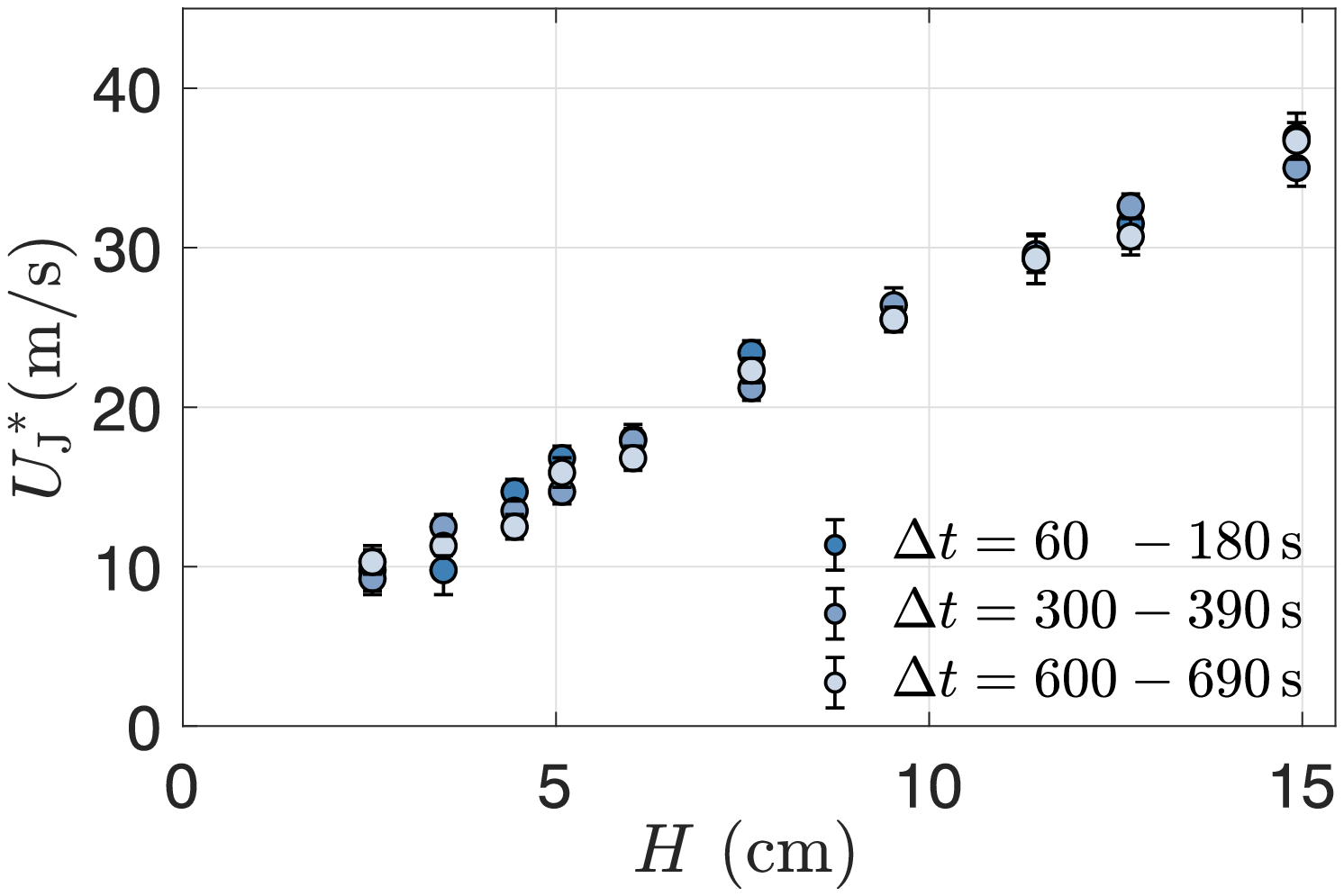}} 
\subfigure[]{\includegraphics[width = 0.42\textwidth]{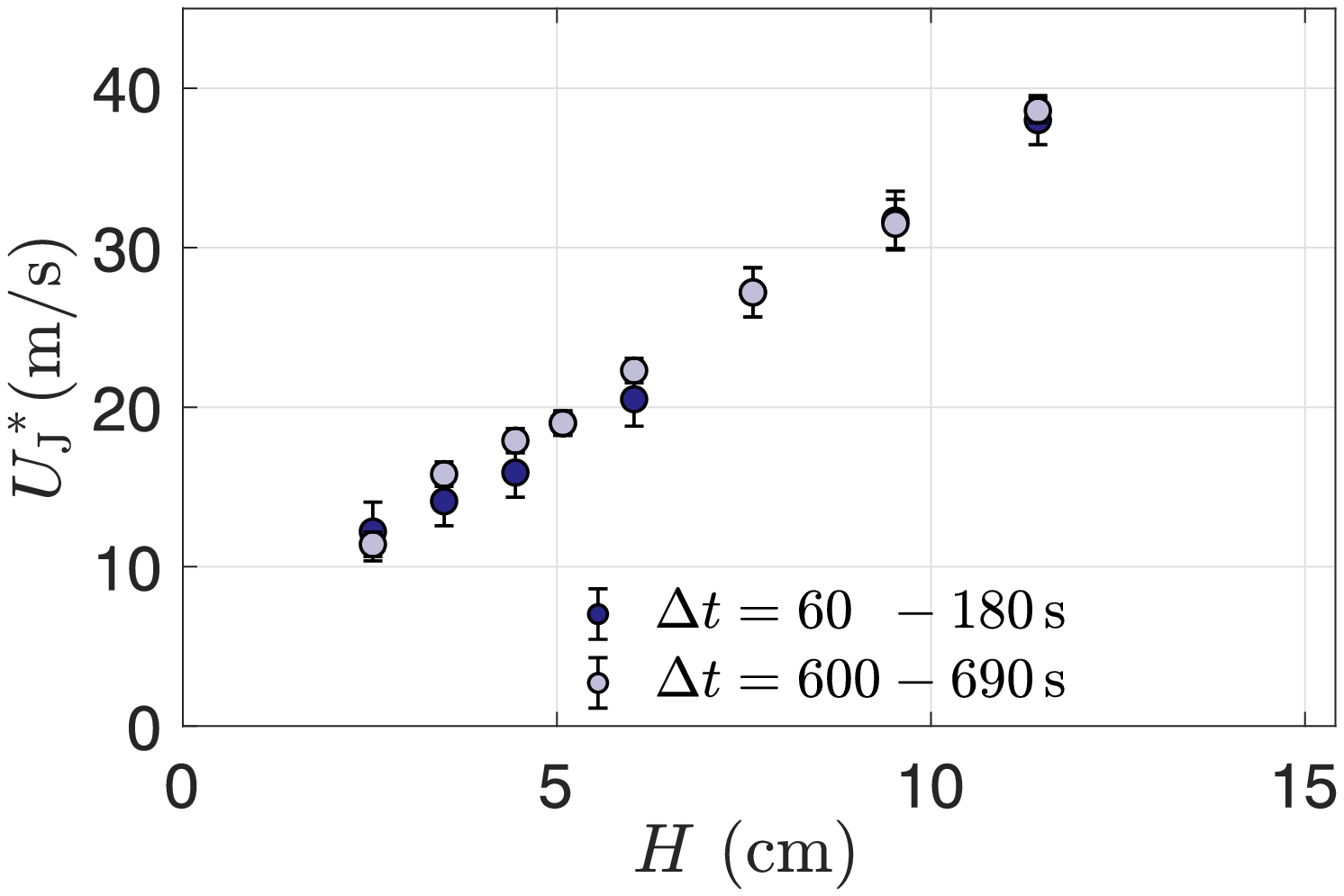}} 
\caption{(a) Schematic of the experimental setup. A turbulent jet exits the nozzle of inner diameter $D$ at the mean velocity $U_{\rm J}$ and impinges the cohesive granular bed placed at a distance $H$. 
(b) Evolution of the cohesive force $F_{\rm c}$ between two particles of diameter $d=800\,\mu{\rm m}$ as a function of the mean coating layer $b$. The circles are the experimental measurements, the line is equation (\ref{eq:cohesive_force}) with $B=230\,{\rm nm}$, and the right axis gives the corresponding value of the cohesion number \cite{gans2020cohesion}. Inset: Schematic showing a glass bead of diameter $d$, and a PBS coating of thickness $b$. (c)-(d) Mean velocity threshold of the jet at the outlet of the nozzle $U_{\rm J}^*$ as a function of the distance between the nozzle and the granular bed $H$ for glass beads of diameter $d = 800 \,\mu{\rm m}$ and two cohesion numbers: (c) ${\rm Co}=4.4$ corresponding to $b=88.5\,{\rm nm}$, and (d) ${\rm Co}=8.0$ corresponding to $b=200\,{\rm nm}$. The different colors indicate different waiting time, between one minute and ten minutes, before measuring the erosion threshold.} 
\label{fig:Figure1_Setup}
\end{figure}

The experimental system used to characterize the erosion of the cohesion-controlled granular material (CCGM) is shown in figure \ref{fig:Figure1_Setup}(a). The particles entirely fill a cylindrical container of diameter 200 mm and height 50 mm. The surface of the granular bed is flattened before each experiment. A metal nozzle with an internal diameter of $D=4.8\,{\rm mm}$ and a length of 50.8 mm (McMaster-Carr) is centered at the vertical above the granular bed. The nozzle is connected to compressed air via a PVC flexible tube. The experiments are carried out at room temperature ($22^{\rm o}{\rm C} \pm 2^{\rm o}{\rm C}$) such that the air has a density $\rho_{\rm a} = 1.19\,{\rm kg\,m^{-3}}$, a dynamic viscosity $\eta_{\rm a} = 1.8 \times 10^{-5}\,{\rm Pa\,s}$, and thus a kinematic viscosity $\nu_{\rm a}=1.5 \times 10^{-5}\,{\rm m^2\,s^{-1}}$. The distance between the outlet of the nozzle and the surface of the granular bed is varied in the range $1\,{\rm cm}<H< 22\,{\rm cm}$. The tubing is connected to a valve to adjust the flow rate $Q_{\rm J}$ of the jet, which is measured with a flowmeter (VWR). The flow rate is varied in the range $4 \times 10^{-5}\,{\rm m^3\,s^{-1}} <Q_{\rm J}< 10^{-3}\,{\rm m^3\,s^{-1}}$, leading to an average velocity of the jet at the outlet of the nozzle of $U_{\rm J} = 4\,Q_{\rm J} /(\pi\,D^2) \in [2,\, 55] {\rm m\,s^{-1}}$, measured with an accuracy of $\pm 2\%$. This leads to a range of jet Reynolds number ${\rm Re}_{\rm J} =U_{\rm J}\,D/\nu_{\rm a} \in [700, \, 18000]$ at the outlet of the nozzle.

The cohesive grains are produced using the method recently developed by Gans \textit{et al.} \cite{gans2020cohesion}. In summary, spherical glass beads (Interactivia) of density $\rho_{\rm g} = 2500 \,{\rm kg \, m^{-3}}$ are sieved to reduce their size distribution, leading to a mean diameter $d = 800 \pm 75\,\mu{\rm m}$. The glass beads are coated by polyborosiloxane (PBS) made from an -OH terminated polydimethylsiloxane (PDMS, Sigma-Aldrich) cross-linked with boric acid (H$_3$BO$_3$, Sigma-Aldrich). Details on the preparation and characterization are provided in Ref. \cite{gans2020cohesion}. The coating thickness $b$ controls the strength of the cohesive force between the particles. The cohesive force is captured by the expression
\begin{equation} \label{eq:cohesive_force}
F_{\rm c}=\frac{3}{2} \pi \gamma d\left(1-e^{-b / B}\right),
\end{equation}
where the characteristic length, $B\approx 230{\,{\rm nm}}$, and the interfacial tension of the PBS, $\gamma=24\,{\rm mN\,m^{-1}}$, are two fitting parameters obtained by measuring the inter-particle cohesive force between two particles \cite{gans2020cohesion}. In this study, the coating thickness $b$ is varied between $0\,{\rm nm}$ (cohesionless grains) and $300\,{\rm nm}$ leading to a cohesive force in the range $0\,\mu{\rm N} \leq F_{\rm c}\leq 66\,\mu{\rm N}$, as shown in figure \ref{fig:Figure1_Setup}(b). The ratio of the cohesive force $F_{\rm c}$ to the apparent weight of a particle $F_{\rm W}$ defines the cohesion number:
\begin{equation} \label{eq:cohesive_number}
\mathrm{Co}= \frac{F_{\rm c}}{F_{\rm W}} = \frac{9\,\gamma\,\left(1-e^{-b / B}\right)}{\rho_{\rm g}\, g\, d^{2}},
\end{equation}
since $\rho_{\rm g} \gg \rho_{\rm a}$. For the coating thicknesses considered in this study, the cohesion number is in the range $0\leq {\rm Co} \leq 10$, where ${\rm Co}=0$ refers to cohesionless grains ($b=0\,{\rm nm}$).

Gans \textit{et al.} \cite{gans2020cohesion} also reported that for thick coatings (typically of order $b=2\,\mu{\rm m}$) the contact time between the grains can influence the cohesive force in their experiments. Since the influence of the contact time depends on the thickness of the coating, we do not expect significant effects of the contact time in our experiments as the coating used is thin enough. Nevertheless, we characterized this effect by repeating our experiments after waiting for different amounts of time. We show in figures \ref{fig:Figure1_Setup}(c)-(d) the effect of the amount of time we allow a granular bed to rest prior to the start of an experiment. In figure \ref{fig:Figure1_Setup}(c), we show the threshold velocity at the outlet of the nozzle for waiting times between 1 and 10 minutes for $b=88.5\,{\rm nm}$ ($\mathrm{Co=4.4}$). Similarly, for $b=200\,{\rm nm}$ ($\mathrm{Co=8}$), we show the difference between waiting one and ten minutes in figure \ref{fig:Figure1_Setup}(d). In both cases, the threshold velocity $U_{\rm J}^*$ at which the grains erode does not change significantly or systematically for different waiting times. Moreover, the repeatability of the experiments is very good and comparable to the experiments performed with cohesionless grains. 

\subsection{Experimental protocol}

We initially prepare the granular bed by pouring a large quantity of grains into the container. We then horizontally move a squeegee along the diameter of the container so that excess grains are removed. The resulting granular material fills the box and exhibits a flat surface, without any noticeable compaction effects. We then place the nozzle at a distance $H$ from the granular bed. For the experiments devoted to measure the erosion threshold, the compressed air is initially turned on at a low flow rate, well below the erosion threshold. The vertical jet impacts the horizontal surface of the granular bed normally, and we increase the flow rate (and thus the velocity of the jet, $U_{\rm J}$) in small increments until the first grains are eroded. Increasing the velocity at the outlet of the nozzle $U_{\rm J}$ proportionally increases the velocity at the surface of the granular bed, $u_{\ell}$. The erosion threshold based on the velocity at the outlet of the nozzle, $U_{\rm J}^*$, is then determined visually as the average of the last velocity where no erosion is visible and the first velocity where grains are eroded \cite{badr2014erosion}. The uncertainty in the measurement of the erosion threshold is the difference between these two velocities. The super-scripted star indicates the thresholds or critical values for all quantities considered in this study. Note that experimentally, the threshold of erosion is determined through the velocity at the outlet of the nozzle, $U_{\rm J}^*$. We will therefore have to relate $U_{\rm J}^*$ with the local velocity at the surface of the granular bed $u_{\ell}^*$.

The measurements of the crater morphology are performed using a laser profilometer (Micro-Epsilon, LLT2900-100) at a frequency of 25 Hz. We first acquire the surface profile before starting the jet, which is later used as a reference during the processing of the results. A removable horizontal plate is placed between the jet and the granular bed to deviate the jet until we set the desired velocity $U_{\rm J}$. The acquisition of the profiles is started when the plate is removed and is stopped when the crater is in an asymptotic state where no noticeable evolution of the profile is visible.

\section{Erosion Threshold} \label{sec:erosion_threshold}

\subsection{Threshold velocity} \label{subsec:thresholdVelocity}

\begin{figure}
\centering
\subfigure[]{\includegraphics[width = 0.49\textwidth]{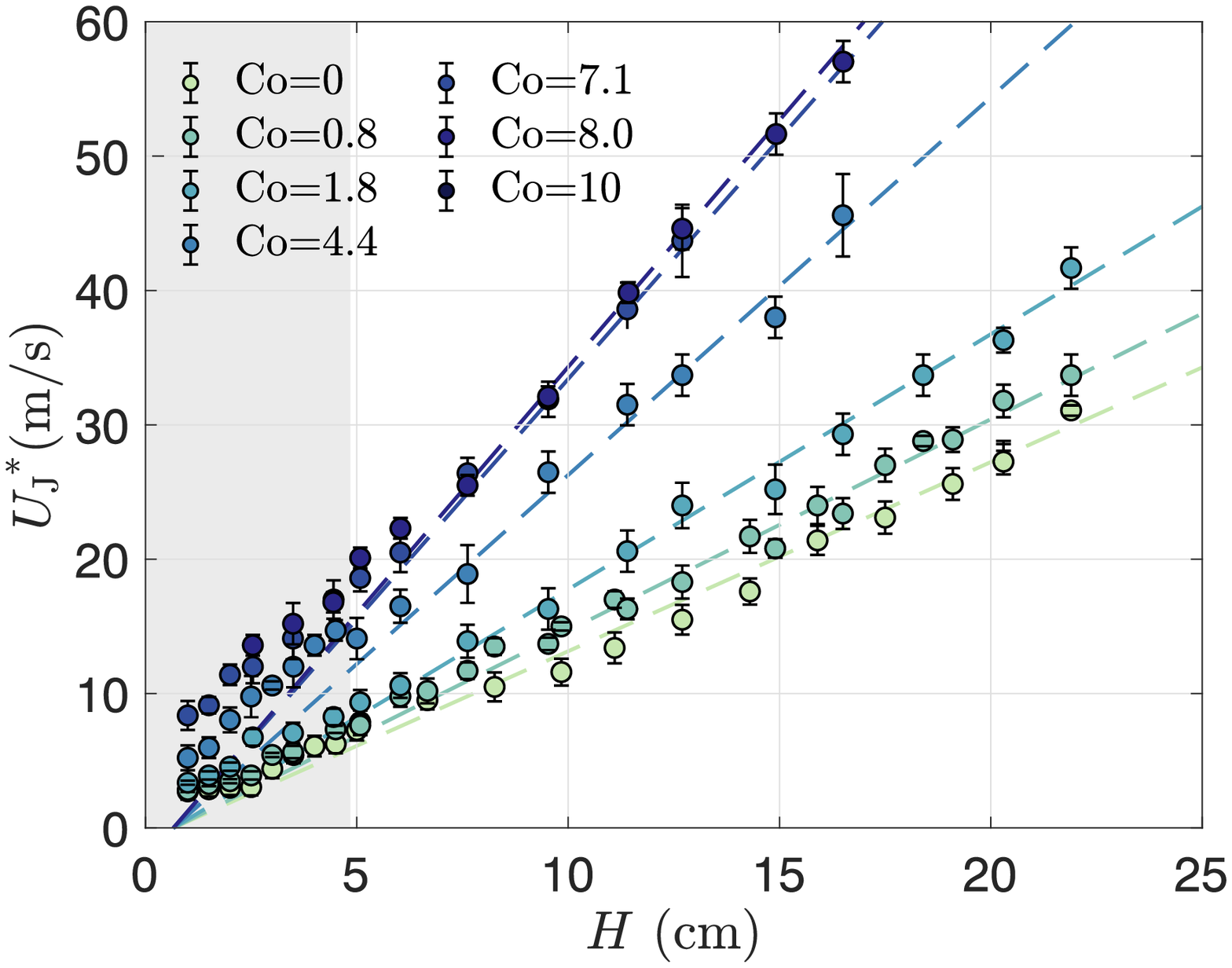}}
\subfigure[]{\includegraphics[width = 0.49\textwidth]{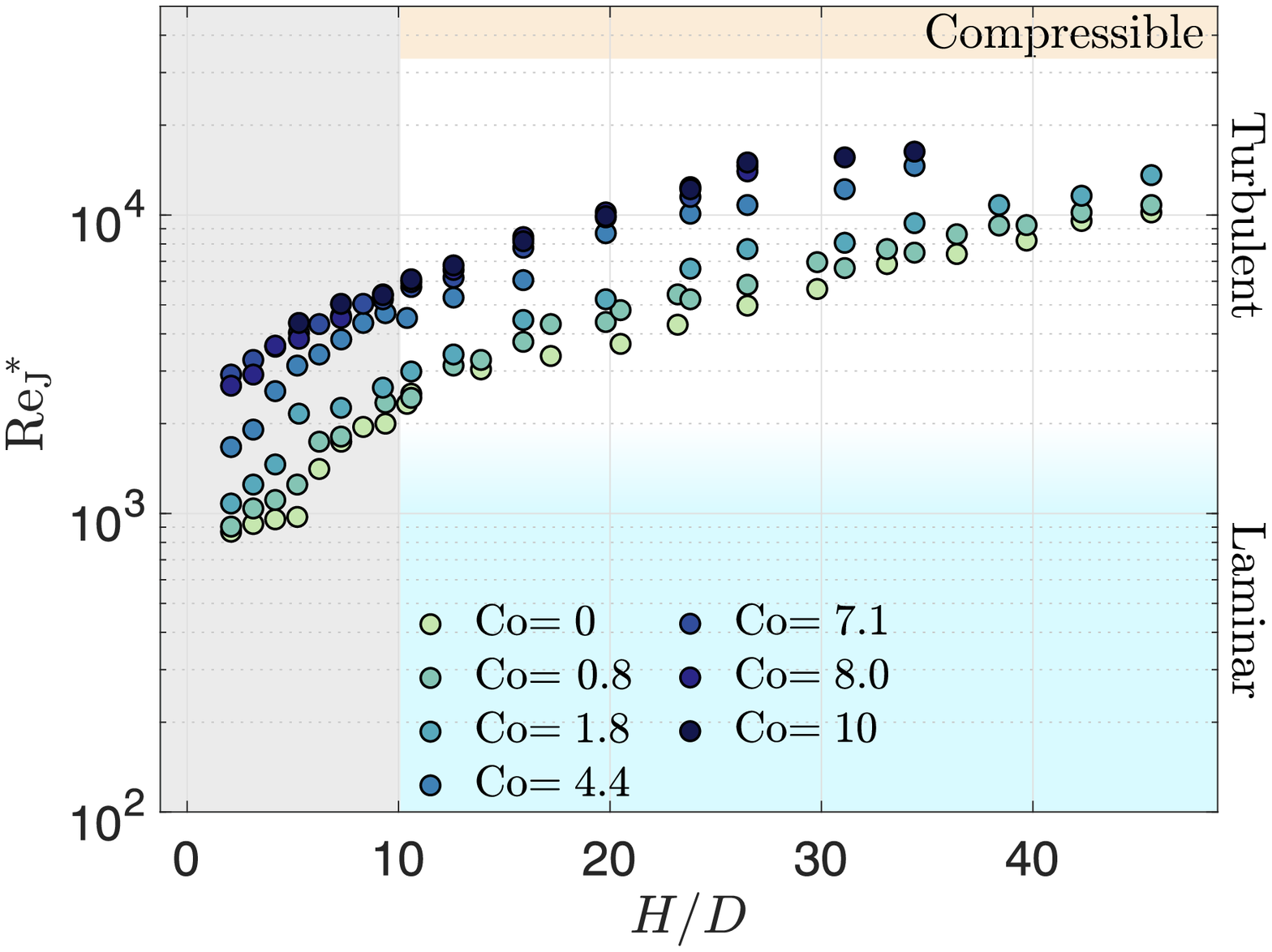}}
\caption{(a) Velocity of the jet at the outlet of the nozzle $U_{\rm J}^*$ at the onset of erosion as a function of the distance between the nozzle and the granular bed $H$ for glass beads of diameter $d = 800 \,\mu{\rm m}$ and different interparticle cohesion from ${\rm Co}=0$ to ${\rm Co}=10$. The dotted lines show the best fit $U_{\rm J}^*=A\,(H+\lambda)$, where the virtual origin is empirically estimated as $\lambda=-0.66\,{\rm cm}$ for all cohesion numbers considered here. (b) Reynolds number associated with the jet ${\rm Re_{\rm J}^*}$ as a function of the dimensionless height, $H/D$, for the threshold values. An estimation of the laminar regime is marked in light blue and the transition to a turbulent jet occurs around ${\rm Re}_{\rm J} \sim 10^3-2\times 10^3$ as shown by the color gradient. Values of ${\rm Re_{\rm J}^*}$ corresponding to compressible jets are shaded in orange. For a fully developed jet ($H/D>10$), all the experiments are in a turbulent-incompressible regime. The gray shaded region ($H/D<10$) in both figures denotes the distance below which the turbulent jet is not considered to be fully developed.}
\label{fig:Figure2_JetVelocityThreshold}
\end{figure}

We first consider the erosion of a granular bed made of cohesionless grains (${\rm Co}=0$) to establish comparisons for experiments with cohesive grains. Figure \ref{fig:Figure2_JetVelocityThreshold}(a) reports the evolution of the critical velocity $U_{\rm J}^*$ beyond which the grains are eroded when varying the distance $H$ between the outlet of the nozzle and the surface of the granular bed. As expected, the larger the distance $H$ is, the larger the velocity of the jet $U_{\rm J}^*$ needs to be for erosion to occur. 
Indeed, for a fixed grain size, the erosion starts when the maximum local velocity at the surface of the granular bed $u_{\rm s,\, max}$ reaches a threshold value, ${u_{\rm s,\, max}}^*$. This threshold local velocity remains constant for any distance of the jet to the granular bed as it only depends on the properties of the granular bed. Since the velocity decreases with the distance to the nozzle, increasing the nozzle to granular bed distance requires increasing the velocity at the outlet of the nozzle, $U_{\rm J}$, to reach the same maximum surface velocity, $u_{\rm s,\, max}$. 

Overall, the behavior observed here with cohesionless grains (light-green circles in figure \ref{fig:Figure2_JetVelocityThreshold}(a)) is consistent with the observation of Badr \textit{et al.} \cite{badr2016crater}. In particular, we observe a plateau value for the threshold velocity $U_{\rm J}^*$ below a distance $H$ of order 3 to 5 cm, corresponding here to $H/D$ of order 5 to 10. Indeed, below this distance the jet may not be fully turbulent yet since the associated Reynolds number ${\rm Re} \sim 1000$ is in the transition range. Besides, even if the jet is already turbulent due to the experimental conditions, it is not yet fully developed for this distance \cite{pope2000turbulent,kwon2005reynolds}. More specifically, an axisymmetric turbulent jet is fully developed when the time-averaged velocity and pressure profiles only depend on the radial position, local width, and the centerline velocity \cite{davidson2015turbulence,pope2000turbulent}. {This occurs} for a dimensionless nozzle-to-bed distance $H/D$ greater than some threshold value. Although there is no theoretical value for this threshold, different values between $H/D=8$ to $H/D=30$ have been reported in past studies \cite{059.Phares-2000,davidson2015turbulence,pope2000turbulent}. Here, we find that for $H/D \gtrsim 10$ (corresponding to $H \gtrsim 4.8 \,{\rm cm}$) our experiments show a monotonic behavior, suggesting that the jet is turbulent and fully developed. A grey shaded region highlights the regime before the jet is fully developed for all plots and where the model developed later will not be valid. For all experiments reporting the threshold velocity, the Reynolds number associated with the jet, ${\rm Re_{\rm J}^*} =U_{\rm J}^*\,D/\nu_a$, is shown in figure \ref{fig:Figure2_JetVelocityThreshold}(b) as a function of the dimensionless distance $H/D$. For a fully formed jet ($H/D \geq 10$), ${\rm Re_J}$ is larger than $2 \times 10^3$ for all our experiments and the jet is thus turbulent. We can also estimate the Mach number for the jet, $\mathrm{Ma_{\rm J}}=U_{\rm J}/c_{\rm a}$, where $c_{\rm a} = 343\, {\rm m\,s^{-1}}$ is the speed of sound in air. This gives us a range $0.006 < \mathrm{Ma_{\rm J}} < 0.160$ here. For all the erosion thresholds the Mach number is less than $0.3$, so that we can consider the jet to be incompressible \cite{059.Phares-2000}. Put together, this illustrates that our experiments are in a turbulent-incompressible regime.

Figure \ref{fig:Figure2_JetVelocityThreshold}(a) also shows the evolution of the threshold velocity $U_{\rm J}^*$ when varying the interparticle cohesion and the distance to the granular bed. Irrespective of the cohesive force and thus the cohesion number ${\rm Co}$ considered here, the threshold velocity of the jet at the outlet of the nozzle increases when increasing the distance to the granular bed. This observation is similar to what we reported above for cohesionless grains. We also observe that for a given distance $H$ to the granular bed, the threshold velocity of the jet $U_{\rm J}^*$ increases with the cohesion between the grains. For example, when the nozzle is located at a distance $H=10\,{\rm cm}$ from the granular bed, the velocity of the jet required to erode the strongly cohesive grains (${\rm Co}=10$) is between 2 and 3 times larger than non-cohesive grains (${\rm Co}=0$). Nevertheless, the global trend is similar with and without cohesion, except for a small difference when the nozzle is close to the granular bed, and the plateau value seems to disappear for cohesive grains. This observation is likely due to the fact that the associated Reynolds number is larger for cohesive grains at a given distance to the granular bed.

To account for the increase in the threshold jet velocity $U_{\rm J}^*$ with the distance to the granular bed $H$ we need to consider the structure of the turbulent jet. This complex configuration has been the topic of various studies. We recall in appendix A the main properties of a turbulent jet impinging on a solid surface that we shall used in the following. In summary, increasing the distance to the surface of the granular bed while keeping $U_{\rm J}$ constant reduces the velocity at the surface. Phares \textit{et al.} An estimate of the centerline velocity at a distance $z=H$ from the outlet of the nozzle of inner diameter $D$ is given by (see Appendix A):
\begin{equation}
U_c (z=H, r=0) = U_{\rm J} \frac{\kappa^{1/2}}{\varepsilon_o} \left(\frac{H + \lambda}{D}\right)^{-1},
\end{equation}
where $\kappa$ and $\varepsilon_o$ are constants associated with the turbulent jet and $\lambda$ is a virtual origin of the turbulent jet. Indeed, Badr \textit{et al.} \cite{badr2016crater} took into account a virtual origin and provided a similar expression for the axial velocity at a dimensionless distance $H/D$: $U_{c}= U_{\rm J}\ K \ (H/D\ +\ \lambda/D)^{-1}$. In this case, $K$ is a numerical prefactor that captures the details of the structure of the jet and depends on the experimental conditions. In this expression, $\lambda/D$ is the dimensionless distance of the nozzle outlet to the virtual origin of the turbulent jet \cite{kotsovinos1976note}. Note that the exact structure and velocity associated with a turbulent jet depend on many parameters such as the precise shape and length of the outlet, the stability, or the air source. It is therefore challenging to obtain quantitative values. As a result, an estimate of the maximum local surface velocity at the onset of erosion (see appendix A) is thus given as:
\begin{equation} 
u_{\rm s,\, max} \simeq U_{c}(z=H, r=0) = K\ U_J \ \left(\frac{H + \lambda}{D} \right)^{-1}.
\label{eq:localVelocity}
\end{equation}

It is important to note that turbulent jets are very sensitive to the experimental conditions and to external perturbations. As a consequence, experimental turbulent jets slightly differ from theoretical models. For the same range of Reynolds number, the values of the decay coefficient $K$  and the position of the virtual origin $\lambda$ can be quite different from one experimental setup to another (see \textit{e.g.}, \cite{chua1998measurements} and references therein). The value of $K$ has been estimated in various studies and presents a significant spread between experiments. For instance, Ref. \cite{059.Phares-2000} report a value of $6.2$ while Ref. \cite{badr2016crater} reports it to be of order 4. Therefore, we consider here $K$ and $\lambda$ as empirical parameters obtained from the experiments with cohesionless grains.

For a given particle size and cohesive force, the erosion should occur at a constant local velocity at the surface of the granular bed ${u_{\rm s,\, max}}^*$.  When increasing the distance of the nozzle to the granular bed, to keep the same value of ${u_{\rm s,\, max}}^*$, one needs to increase the velocity at the outlet of the nozzle such that:
\begin{equation}\label{eq:velocity}
U_{\rm J}^*={{u_{\rm s,\, max}}^*}\,\frac{H+\lambda}{K\,D}.
\end{equation}
Therefore, $U_{\rm J}^*$ should increase as $(H+\lambda)$ with a prefactor $A={u_{\rm s,\, max}}^*/({K\,D})$. We fit the erosion threshold measured for the different cohesions for $H/D > 10$ and reported the results in figure \ref{fig:Figure2_JetVelocityThreshold}(a). We observe that for all cohesive forces, this evolution captures well the increase in $U_{\rm J}^*$ with $H$, when taking $\lambda=-0.66\,{\rm cm}$. The location of the virtual origin in the present case, $\lambda/D=-1.4$, is compatible with values reported in the literature \cite{malmstrom1997centreline,chua1998measurements,badr2016crater}. The prefactor $A$ captures the local erosion velocity as well as the structure of the turbulent jet.

In summary, for each cohesion number, the increase in $U_{\rm J}^*$ with $H/D$ follows the scaling law given by equation (\ref{eq:velocity}) based on the structure of an axisymmetric turbulent jet. The shift between the different curves is due to interparticle cohesion, which is the primary consideration in the following.

\subsection{Jet Shields number} \label{subsec:cohesiveShields}

\begin{figure}
\centering
\subfigure[]{\includegraphics[width = 0.49\textwidth]{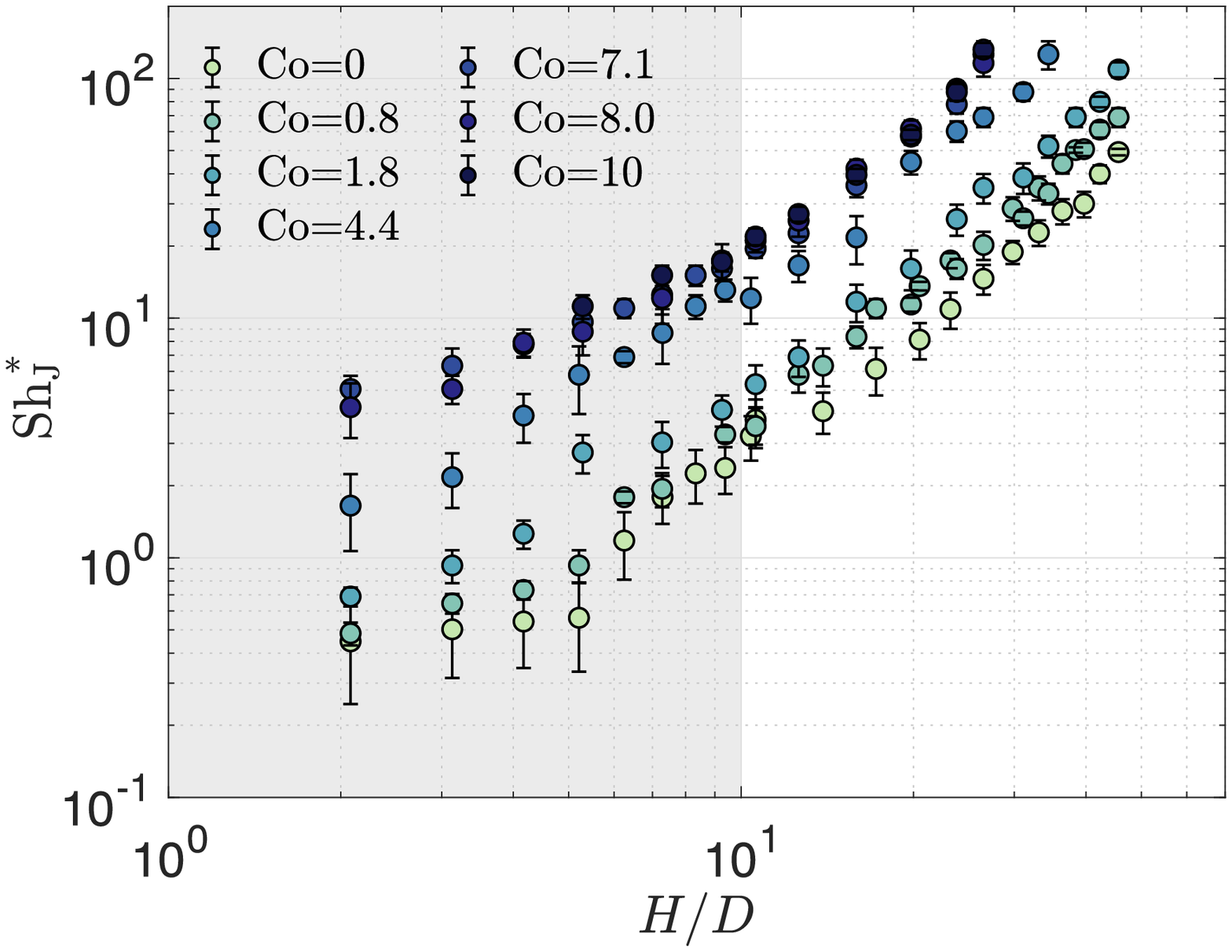}}
\subfigure[]{\includegraphics[width = 0.49\textwidth]{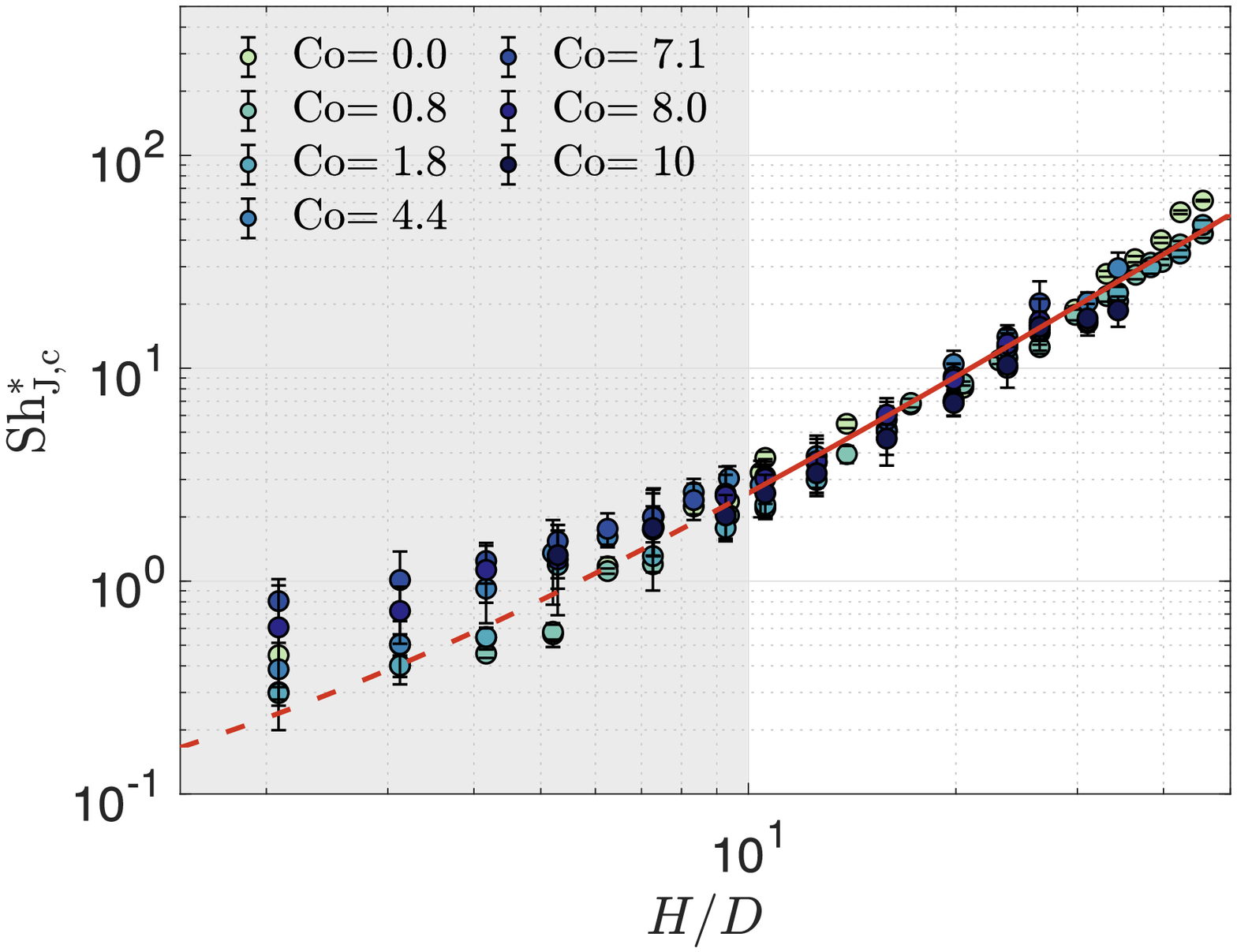}}
\caption{(a) Jet cohesionless Shields number, ${\rm Sh_{\rm J}}^*$, as a function of the rescaled distance $H/D$ and varying cohesion numbers ${\rm Co}$ at the onset of erosion. (b) Jet cohesive Shields number $\mathrm{Sh}_{\mathrm{J}, \mathrm{c}}^*$ given by equation (\ref{eq:cohesive}) with $\alpha= 0.75$ as a function of $H/D$ for the experiments reported in figure \ref{fig:Figure3_ShieldsJet}(a). The solid line corresponds to equation $\mathrm{Sh_{J, c}^*} = C \ (H/D + \lambda/D)^{2}$ with $C = 0.02$ and $\lambda/D=-1.4$. In both figures the grey region denotes the distance below which the turbulent jet is not fully developed ($H/D <10$).}
\label{fig:Figure3_ShieldsJet}
\end{figure}

All experiments performed here are for Reynolds numbers ${\rm Re_J} \geq 10^3$ and therefore only involve turbulent jets. Following the approach of Refs. \cite{badr2014erosion,brunier2017erosion}, we first consider an inertial stress based on the velocity at the outlet of the nozzle, $\tau_{\rm f} =\rho_{\rm a}\, {U_{\rm J}}^{2}$, to define a cohesionless jet Shields number :
\begin{equation}
{\rm Sh_{\rm J}}=\frac{\rho_{ \rm a}\, {U_{\rm J}}^{2}}{(\rho_{\rm g}-\rho_{\rm a})\, g\, d}.
\label{eq:Shields}
\end{equation}
The evolution of $\rm Sh_{\rm J}^*$ as a function of the dimensionless distance to the granular bed $H/D$ is reported in figure \ref{fig:Figure3_ShieldsJet}(a). The threshold jet Shields number $\rm Sh_{\rm J}^*$ exhibits the same trend for the range of cohesion numbers studied here but are shifted. Indeed, the larger the cohesion number ${\rm Co}$ is, the larger $\rm Sh_{\rm J}^*$. This result is expected since the Shields number is the ratio of the forces eroding the grain, \textit{i.e.}, the drag, and the forces stabilizing it. The weight of the grain is stabilizing the grain on the granular bed, but when a cohesive force between particles is added, one now needs to account for this additional stabilizing cohesive force $F_{\rm c}$ in the Shields number. Thus, to reach the critical value of the local Shields number marking the onset of erosion of the grains, the destabilizing force must increase. The velocity at the outlet of the nozzle must be correspondingly larger. In summary, a jet Shields number based only on the gravitational force and the inertial drag at the outlet of the nozzle does not capture this difference. 

For a cohesive granular medium, the destabilizing force remains the turbulent drag force acting on the grain, and the total stabilizing force, now comes from the gravity and cohesive forces, $F_{\mathrm{W}}$ and $F_{\mathrm{c, \text{tot}}}$, respectively. Here, $F_{\mathrm{c, \text{tot}}}$ is the cohesive force coming from the contact between a grain and its neighbors. For cohesionless grains, the weight of the grain leads to a friction force that prevents the grains to be eroded \cite{andreotti2013granular}. In the case of the model cohesive grains used here, the friction coefficient does not seem to be modified by the coating layer \cite{gans2020cohesion}. Nevertheless, the presence of cohesive forces leads to a more challenging condition to describe. Indeed, the cohesive forces leads to additional vertical components (in the direction of the gravity), as well as horizontal components. Assuming a homogeneous cohesion, the resulting cohesive force can be estimated as a sum of the different components of the cohesive force between the particles so that $F_{c, \text {tot}}= \sum_{i=1}^{N} F_{{\rm c},i}$. where $F_{{\rm c},i}$ is the resulting cohesive force between two particles in contact and $N$ is the number of neighboring grains in contact. Note that the total cohesive force will depend on the angle between the neighboring grains and the direction of the drag force on the eroding particle. Besides, we have assumed here for the sake of simplicity, that the erosion occurs at the scale of an individual grain. However, the erosion of clusters of grains could also play a significant role, as we shall discuss later. To account for the total contribution of the cohesive effects, we introduce an empirical coefficient $\alpha$ accounting for the number of cohesive contacts and the orientation of a grain with respect to the drag force. As a result, the total cohesive force acting on a grain is written as $F_{c, \text {tot}}=\alpha\, F_{{\rm c}}$. Assuming a classical disposition of the grains at the surface (tetrahedral, pyramidal, etc.), $\alpha$ can be estimated to be of order 1 \cite{agudo2012incipient}. Using the expression of the cohesive force between two grains $F_{\rm c}$, the jet cohesive Shields number can be defined as: 
\begin{equation}
\mathrm{Sh}_{\mathrm{J}, \mathrm{c}}=\frac{F_{\rm J}}{F_{\rm  W}+F_{c, \text{tot}}}=\frac{\mathrm{Sh}_{\mathrm{J}}}{1+\alpha\,F_{\rm  c} / F_{\rm W}}.
\end{equation}
Further, using the definition of the cohesion number ${\rm Co}$, we obtain a jet cohesive Shields number:
\begin{equation}
\mathrm{Sh}_{\mathrm{J}, \mathrm{c}}=\frac{\mathrm{Sh}_{\mathrm{J}}}{1+\alpha \,\mathrm{Co}}=\frac{1}{{1+\alpha \mathrm{Co}}}\,\frac{\rho_a\, {U_{\rm J}}^{2}}{(\rho_{\rm g}-\rho_{\rm a})\, g\, d}.
\label{eq:cohesive}
\end{equation}
Note that a similar definition of the cohesive Shields number was also suggested in other configurations or flow conditions (see \textit{e.g.}, \cite{ternat2008erosion,andreotti2013granular,brunier2020generalized}). Figure \ref{fig:Figure3_ShieldsJet}(b) shows the rescaling of the experimental data reported previously using this factor accounting for the cohesion. We observe that for all cohesion, we can collapse the results on a master curve for $\alpha = 0.75$, in good agreement with equation (\ref{eq:cohesive}). The effect of the cohesive forces is captured by the empirical factor $\alpha$, which depends on the geometry and the cohesion number ${\rm Co}$. Erosion occurs when the cohesive Shields number is larger than the threshold Shields number, \textit{i.e.} $\mathrm{Sh}_{\mathrm{J}, \mathrm{c}} \gtrsim {\mathrm{Sh}_{\mathrm{J}, \mathrm{c}}}^*$. The threshold value depends on the properties of the grains, cohesion between them as well as the nature and geometry of the erosive stresses. We should emphasize that the definition of a cohesive Shields number to account for inter-particle cohesion was suggested in Refs. \cite{ternat2008erosion,andreotti2013granular}, and also considered by Brunier-Coulin \textit{et al.} \cite{brunier2020generalized} who obtained a similar relation. The value of $\alpha$ reported from their experiments performed at larger cohesion number is slightly larger that the values estimated here \cite{brunier2020generalized}. The difference may come from the definition of the cohesive force, which is considered here at the particle scale whereas it was measured in a macroscopic sample in their experiments for the beads considered in their erosion experiments. In addition, the range of cohesion number here is much smaller than in Ref. \cite{brunier2020generalized} so that the erosion could occur differently. Nevertheless, the difference in the empirical values of $\alpha$ is small, so that it suggests that such a definition of the cohesive Shields number is valid both for reversible and irreversible cohesion.

\subsection{Local Shields number} \label{subsec:localShields}

\begin{figure}
\centering
\subfigure[]{\includegraphics[width = 0.49\textwidth]{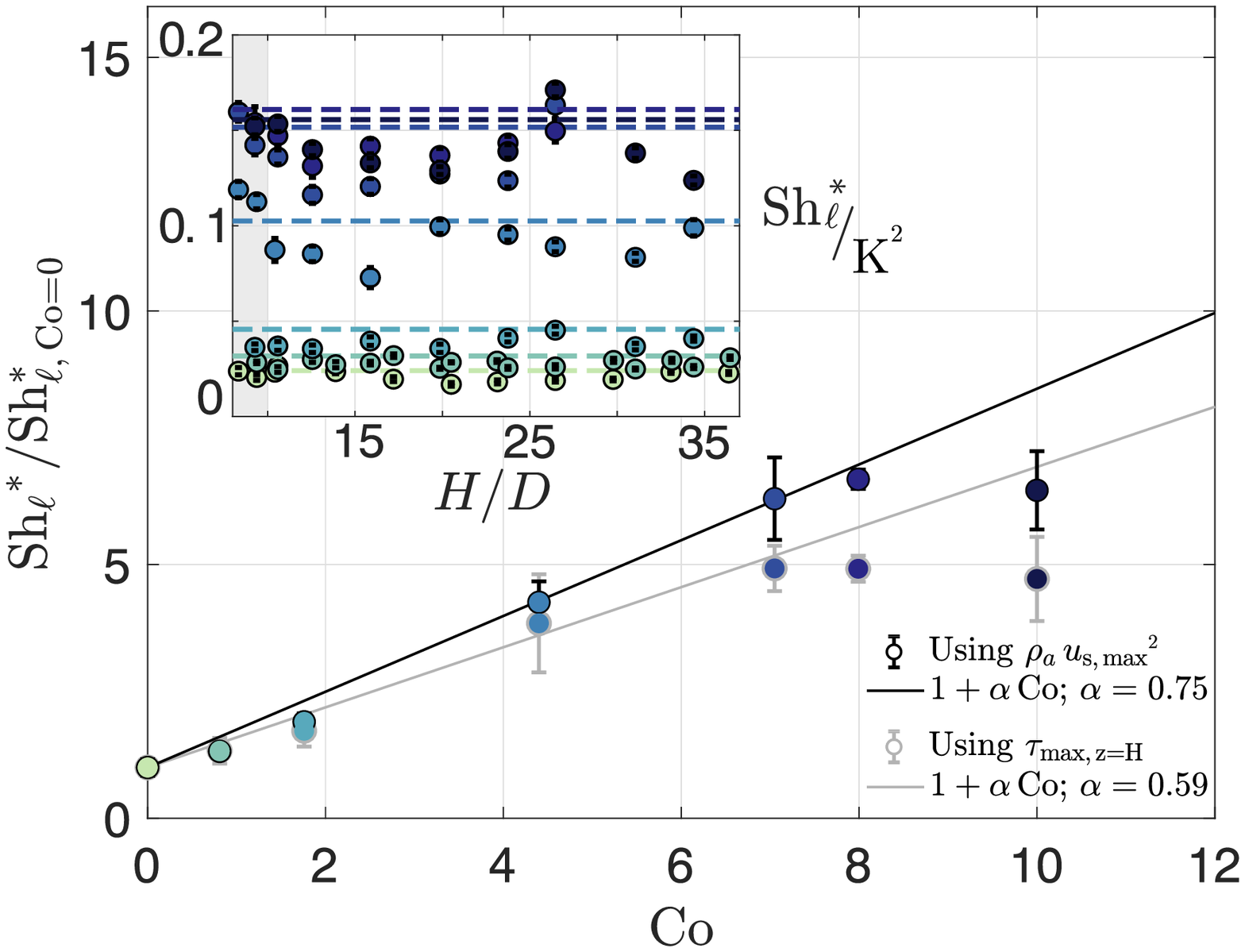}}
\subfigure[]{\includegraphics[width = 0.49\textwidth]{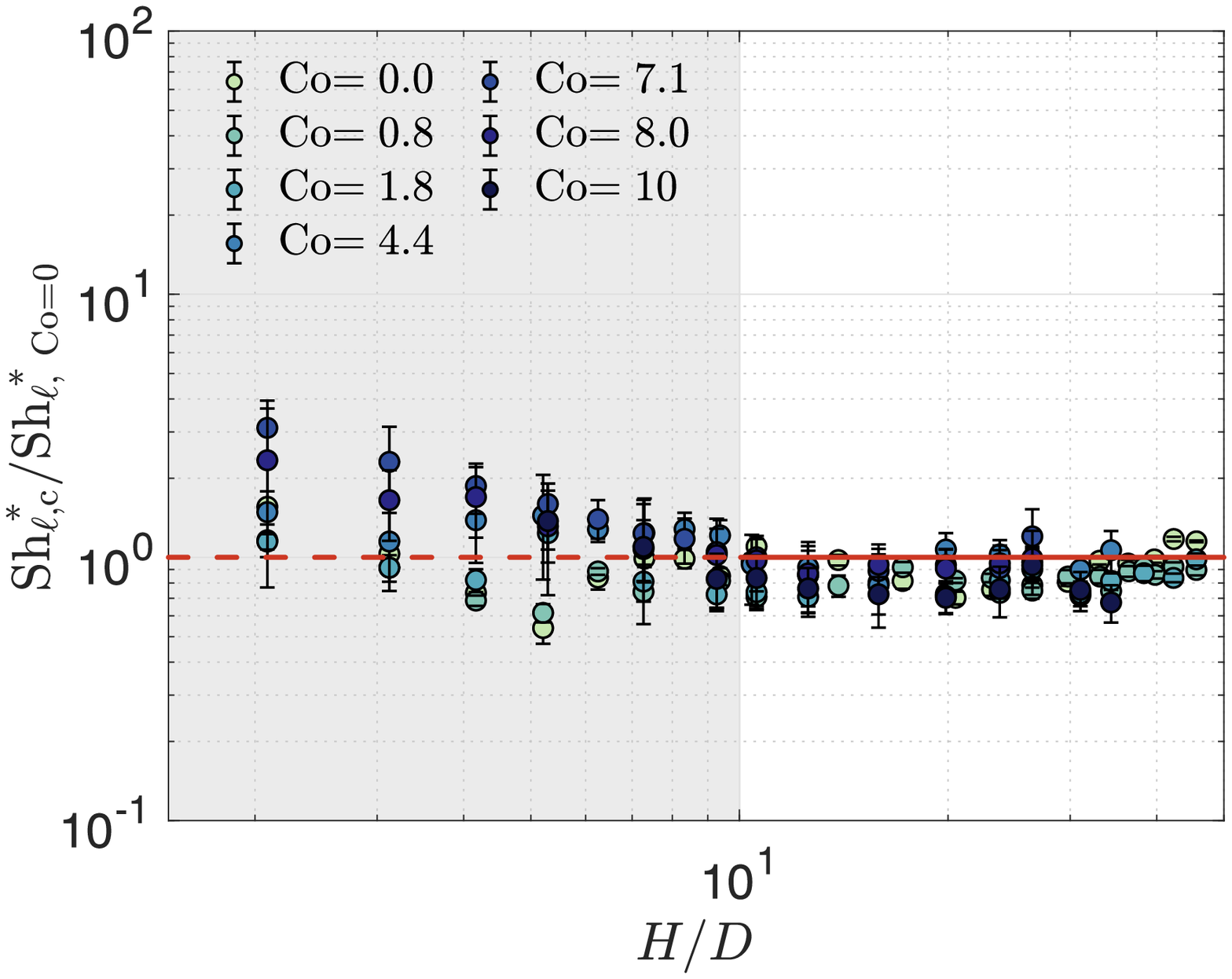}}
\caption{(a) Evolution of the local erosion threshold values for $\mathrm{Sh}_{\ell}^*$ divided by the cohesionless threshold $\mathrm{Sh}_{\ell, \ {\rm Co=0}}^*$ for $H/D \gtrsim 10$ using the maximum surface velocity (in black) and the maximum shear stress on the surface (in grey). The solid lines show a fit of the form $\rm{Sh}_{\ell} / \rm{Sh}_{\rm {\ell, \ Co=0}} = 1+\alpha\ \mathrm{Co}$ for cohesion numbers $0 \leq \mathrm{Co} \leq 8$ with $\alpha=0.75 \pm 0.04$ (R-square: 0.9944) for the maximum local velocity and $\alpha=0.59 \pm 0.07$ (R-square: 0.9653) using the maximum shear stress. Inset: Threshold values for $\mathrm{Sh}_{\ell}^*$ plateau for $H/D \gtrsim 10$ using the maximum surface velocity, and setting $K=1$. The colored dashed lines display the averages for a given $\mathrm{Co}$ number. (b) Local cohesive Shields number, $\mathrm{Sh^*_{\rm{\ell, \, c}}}$ divided by the cohesionless local Shields number $\mathrm{Sh^*_{\rm{\ell, \ Co=0}}}$ when varying the dimensionless distance $H/D$. The gray shaded area shows $H/D \lesssim 10$, \textit{i.e.} where the jet is not fully developed.}
\label{fig:Figure4_ShieldsLocal}
\end{figure}

The characterization of the erosion through a Shields number is usually done at the particle scale, \textit{i.e.}, a local criterion \cite{andreotti2013granular}. Here, the flow configuration (turbulent jet) is more complicated than a simple uni-directional flow parallel to the surface of the granular bed. Nevertheless, an analysis can be performed a the local scale with some fitting parameters to describe the turbulent jet. The particle Reynolds number at the onset of erosion can be expressed as $\mathrm{{Re}_{\rm p}}^* = ({u_{\rm s,\, max}}^*\, d \, \rho_{\rm a}) / \eta_{\rm a}$. This expression can be rewritten using equation (\ref{eq:localVelocity}) as: 
\begin{equation} 
\mathrm{Re}_{\rm p}^*=K\, {U_{\rm J}^*} \left(\frac{d \ \rho_{\rm a}}{\eta_{\rm a}}\right) \left(\frac{H}{D} + \frac{\lambda}{D}\right)^{-1}.
\label{eq:particleReynolds}
\end{equation}
For our fully developed and axisymmetric turbulent jet ($H/D \gtrsim 10$), we find $\mathrm {Re_p} > 25 K$. Using $K\simeq 4$ \cite{badr2016crater} we obtain $\mathrm{Re_p}^* \gtrsim 100$. The relatively large value of $\mathrm{Re_p}$ leads to an almost constant value of $\mathrm{Sh}^*_\ell$ at the onset of erosion \cite{shields1936application,buffington1999legend,guo1997discussion}. We should emphasize that since $K$ is only an estimated value, the value of $\mathrm{Re}_{\rm p}$ is only an order of magnitude.

We can define the ratio between the local fluid forces at the bed surface and the apparent weight acting on a grain. Since we are interested in the onset of erosion, we consider the maximum local flow velocity ${u_{\rm s,\, max}}$ at the surface of the granular bed. Using equation (\ref{eq:localVelocity}), the local Shields number at the particle scale, $\mathrm{Sh_{\ell}}$, can be written as:
\begin{equation}
\mathrm {Sh_{\rm \ell}} = \frac{F_{\rm s}}{F_{\rm W}} = \frac{\mathrm{\tau}_{\rm{s}}} {(\rho_{\rm g}-\rho_{\rm a})\, g\, d} = \frac{\rho_{\rm a}\, {u_{\rm s,\, max}}^{2}}{(\rho_{\rm g}-\rho_{\rm a})\, g\, d} = {K^2}\frac{\rho_a \ { {U_{J}}}^{2}}{(\rho_{\rm g}-\rho_{\rm a})\, g\, d}\ \left(\frac{H}{D} + \frac{\lambda}{D}\right)^{-2}.
\label{eq:localShields}
\end{equation}

For a fully developped jet, the threshold local Shields number plateaus for all $H/D$ [inset of figure \ref{fig:Figure4_ShieldsLocal}(a)]. Since we have not yet taken into account the effect of cohesion, the plateau values are higher for more cohesive grains. Using an identical argument to the previous section, we can account for cohesion with the same correction factor $1/(1 + \alpha\, \mathrm{Co})$ to describe a cohesive local Shields number, $\mathrm{Sh_{\ell, \, c}}$:
\begin{equation} 
{\rm Sh_{\rm \ell, c}}=\frac{F_{\rm s}}{F_{\rm w} + F_{\rm c}} = \frac{\rm{Sh}_{\ell}}{1 +  \alpha {\rm Co}}=K^2 \left(\frac{1}{1 + \alpha {\rm Co}}\right) \frac{\rho_a \ {U_{\rm J}}^{2}}{(\rho_{\rm g}-\rho_{\rm a})\, g\, d}\ \left(\frac{H}{D} + \frac{\lambda}{D}\right)^{-2},
\label{eq:localShields_coh}
\end{equation} 
The average plateau values reported in the inset of figure \ref{fig:Figure4_ShieldsLocal}(a) are normalized by $\mathrm{Sh_{\rm{\ell, \, Co=0}}^*}$ (\textit{i.e.}, the threshold local Shields number for the cohesionless grains) in figure \ref{fig:Figure4_ShieldsLocal}(a). Since we do not experimentally measure the local velocity, we do not have an exact value for the parameter $K$. However, by normalizing by $\mathrm{Sh_{\rm{\ell, \, Co=0}}^*}$, we remove the dependence on the parameter $K$, and can find an experimental value for $\alpha$ using a fit to this trend. For cohesion numbers $0 \leq \mathrm{Co} \leq 8$, the cohesive contribution to the local Shields number can be captured by the expression: $\mathrm{Sh}_{\ell, \, c} / \mathrm{Sh}_{\rm{\ell, \ Co=0}} = (1 + \alpha {\rm Co})$, with $\alpha = 0.75 \pm 0.04$ ($R^2$: 0.9944) as illustrated by the black line in figure \ref{fig:Figure4_ShieldsLocal}(a).

Another possible approach to estimate the local Shields number would be to consider the magnitude of the maximum shear stress on the impinged surface (see \textit{e.g.}, \cite{brunier2020generalized}). At the onset of erosion, we expect the location of the maximum shear stress to roughly corresponds to the location of $u_{\rm s,\, max}$ (see Appendix A). This method leads to results similar to the ones described earlier. Introducing $\tau_{\rm max, \, z=H}$ from equation (\ref{eq:maxShear}) into $\mathrm{\tau}_{\rm{s}}$ in equation (\ref{eq:localShields}) and using a similar factor to account for cohesion, we get:
\begin{equation}
{\rm Sh_{\rm \ell, c}}= \frac{\mathrm{\tau}_{\rm max, \, z=H}} {(\rho_{\rm g}-\rho_{\rm a})\, g\, d} \left(\frac{1}{1 + \alpha {\rm Co}}\right) = \frac{K_2}{{{\rm Re}_{\rm J}}^{1/2}} \, \left(\frac{1}{1 + \alpha {\rm Co}}\right) \frac{\rho_a \, {U_{\rm J}}^2} {(\rho_{\rm g}-\rho_{\rm a})\, g\, d} \, \left(\frac{H}{D} + \frac{\lambda}{D}\right)^{-2},
\label{eq:localShields_coh2}
\end{equation} where $K_2$ is a different experimental constant and ${\rm Re}_{\rm J}$ is the Reynolds number at the outlet of the nozzle. The main difference between these two approaches is a difference in the arrangement of experimental constants. Indeed, the constant $K$ used in the approach with the local velocity is also expected to depend on ${\rm Re}_{\rm J}$. Normalizing by the cohesionless case just as above, we find $\alpha = 0.59 \pm 0.07$ ($R^2$: 0.9653) for $0 \leq \mathrm{Co} \leq 8$ using the maximum shear stress on the surface (figure \ref{fig:Figure4_ShieldsLocal}(a)).

In both cases, the values of $\alpha$ are similar, and slightly smaller than $1$. These approaches show that the Shields number needs to be modified to capture the erosion of cohesive grains as observed in Ref. \cite{brunier2020generalized}. The value of the prefactor found here is slightly smaller than the one reported in Ref. \cite{brunier2020generalized}. However, different reasons could explain this observation. The prefactors are not the same in the definition of the cohesion number Co, the type of cohesion are different (reversible adhesion versus brittle cementation), and the hydraulic regimes are completely different. In addition, Ref. \cite{brunier2020generalized} converted their tensile yield measurements at the bond scale to a larger scale through Rumpf's law, which is known to underestimate the macroscopic stress and so increase the value of $\alpha$. Nevertheless, the values found here and reported in Ref. \cite{brunier2020generalized} are still of the same order of magnitude. This result suggests that the nature of the cohesion does not significantly influence the definition of the cohesive Shields number.

In summary, to describe the erosion of a cohesive granular bed by a turbulent jet one need to introduce a term that captures the decrease of the velocity with the distance, a term that accounts for the cohesion and a prefactor that contains information regarding the turbulent jet. Our experimental results at the onset of erosion are summarized in figure \ref{fig:Figure4_ShieldsLocal}(b), which shows that both the role of the cohesion and the distance of the jet are well captured by equation (\ref{eq:localShields_coh}), using $\alpha = 0.75$.

\section{Beyond the erosion threshold} \label{sec:crater_morphology}

Beyond the erosion threshold, the turbulent jet impinging the granular bed leads to the transport of grains and the formation of a crater \cite{lamarche2015cratering,badr2016crater}. The present situation is different from past studies that have relied on solid bridges and a laminar flow \cite{brunier2016etude,brunier2020generalized}, where once the erosion threshold is reached, such bonds are irreversibly broken. In other words, while cohesion can be used to determine erosion thresholds, the dynamics and the transport that occur after reaching the threshold are not influenced by cohesive effects anymore. In the present case, the cohesion-controlled granular material allows us to systematically study the crater formation when the bonds between grains can break and reform with new grains with the same cohesive force. 

We consider in the following the morphology of the craters and the influence of the three different parameters: the cohesion number $\rm{Co}$, the dimensionless distance to the granular bed $H/D$, and the velocity of the jet $U_{\mathrm J}$. Alternatively, instead of considering $U_{\mathrm J}$ and $H/D$, we can account for both in the Shields number using the expressions for the cohesive Shields number at the jet-scale and at the local scale developed in the previous section. The difference between the Shields number and the threshold Shields number can be used to further characterize the craters observed \cite{badr2016crater}.

\begin{figure}
\centering
\subfigure[]{\includegraphics[width = 0.49\textwidth]{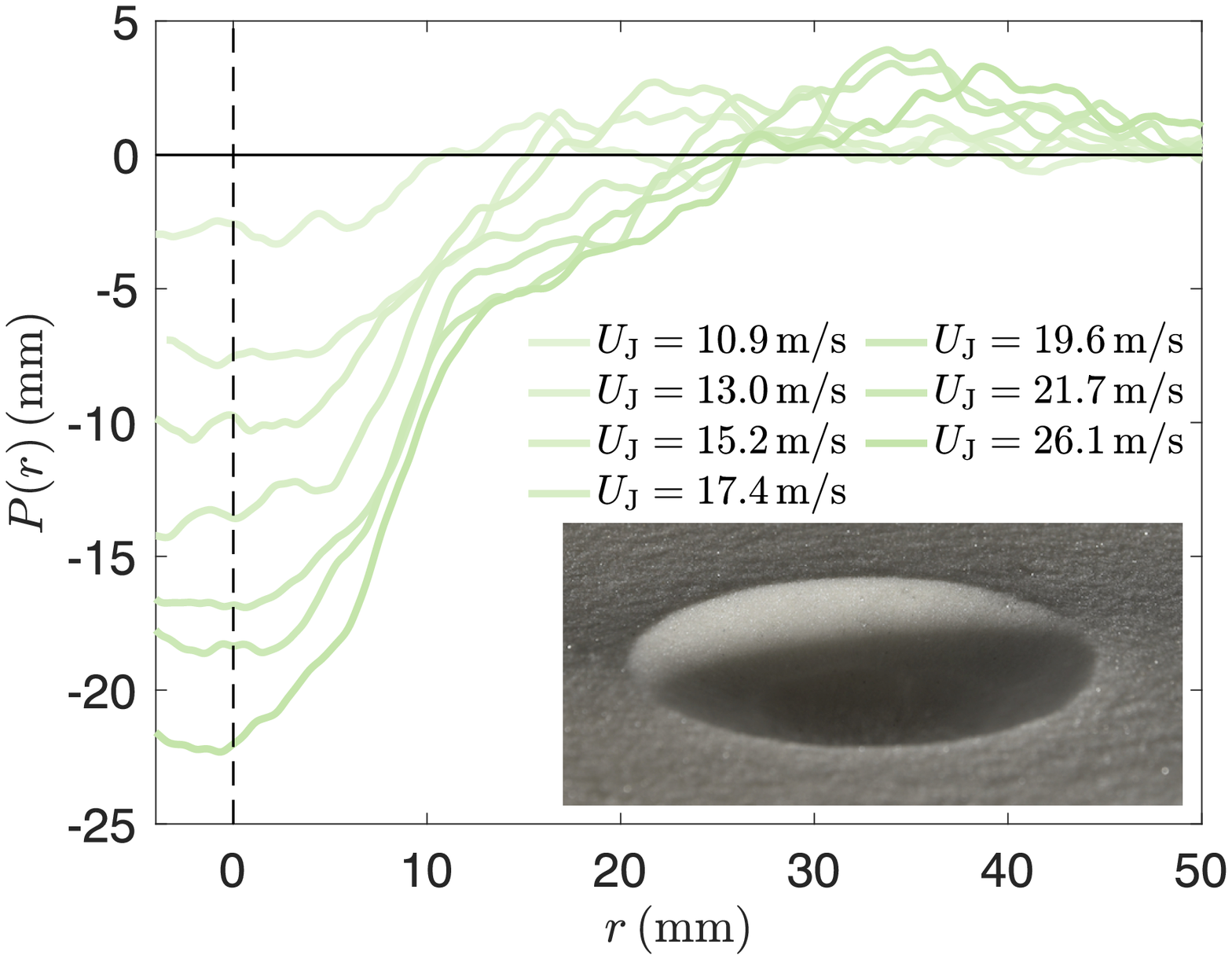}}
\subfigure[]{\includegraphics[width = 0.49\textwidth]{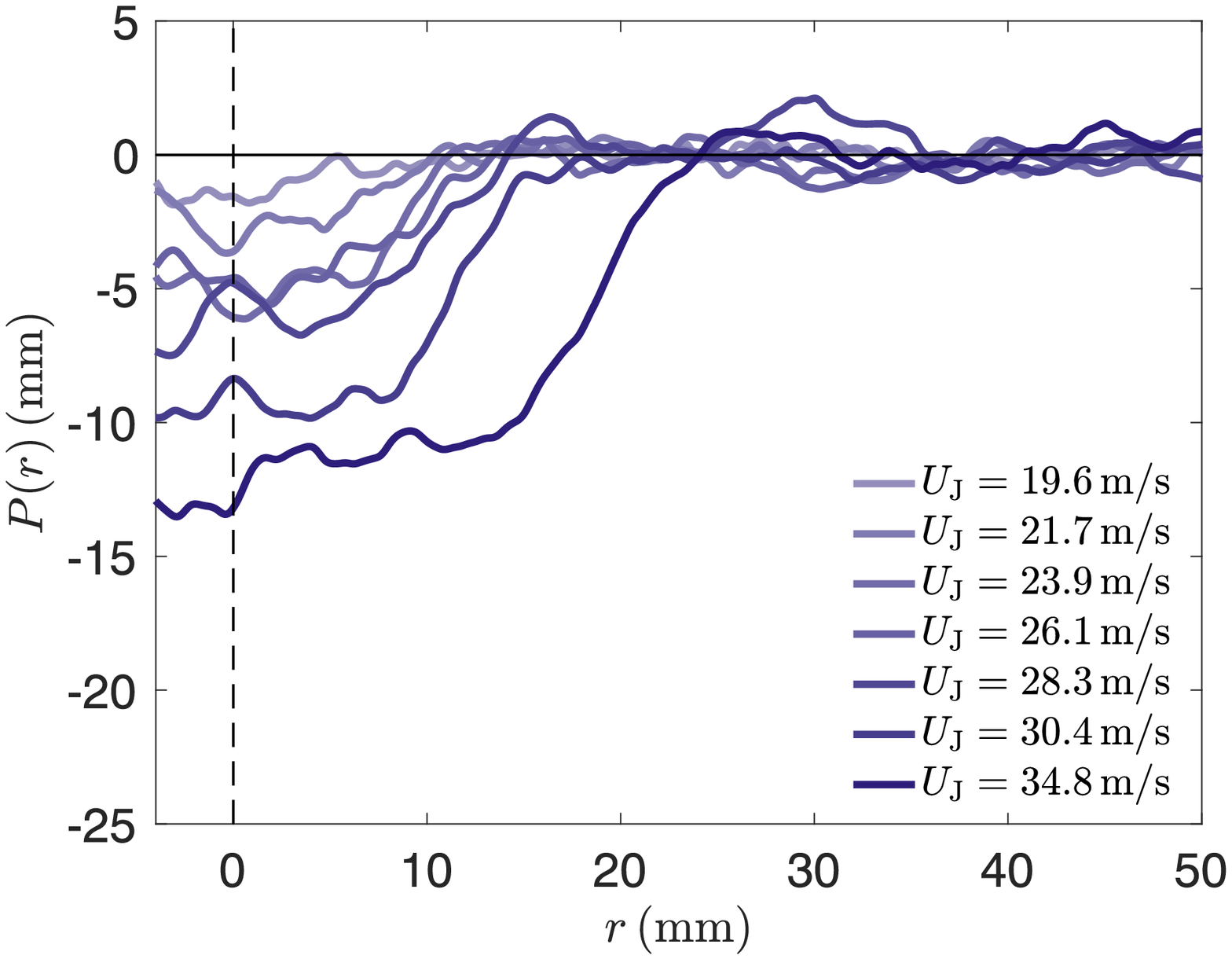}}
\caption{Evolution of the shape of the asymptotic crater for $H=6\,{\rm cm}$ ($H/D=12.6$) when varying the jet velocity $U_{\rm J}$ (a) for cohesionless grains, and (b) for cohesive grains with $\mathrm{Co}=8.0$. The jet velocity $U_{\rm J}$ and color code are indicated in each figure. Darker colors indicate larger velocities at the nozzle for the same $H/D$. The inset in (a) shows a picture of crater observed for cohesionless grains.}
\label{fig:Figure11_App_CraterMorphUj}
\end{figure}

We report in figure \ref{fig:Figure11_App_CraterMorphUj}(a) the evolution of the asymptotic crater morphology for a fixed dimensionless distance $H$ and different jet velocities $U_{\mathrm J} > U_{\mathrm J}^*$ such that the granular bed undergoes erosion and displays an axisymmetric crater. We show here about half of the crater to maximize the resolution of the profilometer. We ensured by direct visualization that the crater is axisymmetric in all our experiments (see inset of figure \ref{fig:Figure11_App_CraterMorphUj}(a)). The asymptotic state refers to the shape of the crater when it stops evolving. In this state, the crater is in dynamic equilibrium due to the radial outflow associated with the jet on the granular crater and the radial inflow of grains due to gravity. We recover the two main types of craters that have been reported for non-cohesive grains \cite{kobus1979flow,aderibigbe1996erosion,mazurek2007scour,lamarche2015cratering}: (i) a weakly deflected regime where the crater is characterized by a parabolic shape (type I), observed here for $U_{\mathrm J}<15\,{\rm  m\,s^{-1}}$  and (ii) a strongly deflected regime where the crater has a parabolic bottom followed by a constant slope due to the accumulation of grains at the edges of the crater (type II), observed here for $U_{\mathrm J} > 15\,{\rm m\,s^{-1}}$. Note also that the evolution of the shape of the crater leads to a feedback that modifies the turbulent jet, which in turns can also modify the crater formed.

The situation with cohesive grains (${\rm Co}=8$) is shown in figure \ref{fig:Figure11_App_CraterMorphUj}(b). In this case, for the same distance to the granular bed, the erosion occurs at larger velocities, in agreement with the previous section. Besides, the evolution of the depth of the crater when increasing the jet velocity seems weaker. A main difference arises in the shape of the crater. Whereas cohesionless grains quickly lead to a type II crater, such geometry is not observed for this cohesion and within the range of velocity $U_{\rm J}$ considered here. Qualitatively, this observation can be explained by the fact that since a stronger flow velocity is required to erode the cohesive grains, the local Shields number needs to be larger. Once a grain is eroded and transported by the turbulent flow, cohesive forces have weaker or no more effects since the grain has less or no contact with other cohesive grains. Therefore, apart from when the particle is bouncing, the cohesion does not modify its dynamics anymore. As a result, the particle can be considered as cohesionless when suspended in the air. The transport rate, which is proportional to the flow velocity, is consequently larger. As a result, the grains are transported further from the edge of the crater and cannot fall back in the crater to lead to a type II crater.

\begin{figure}
\centering
\subfigure[]{\includegraphics[width = 0.49\textwidth]{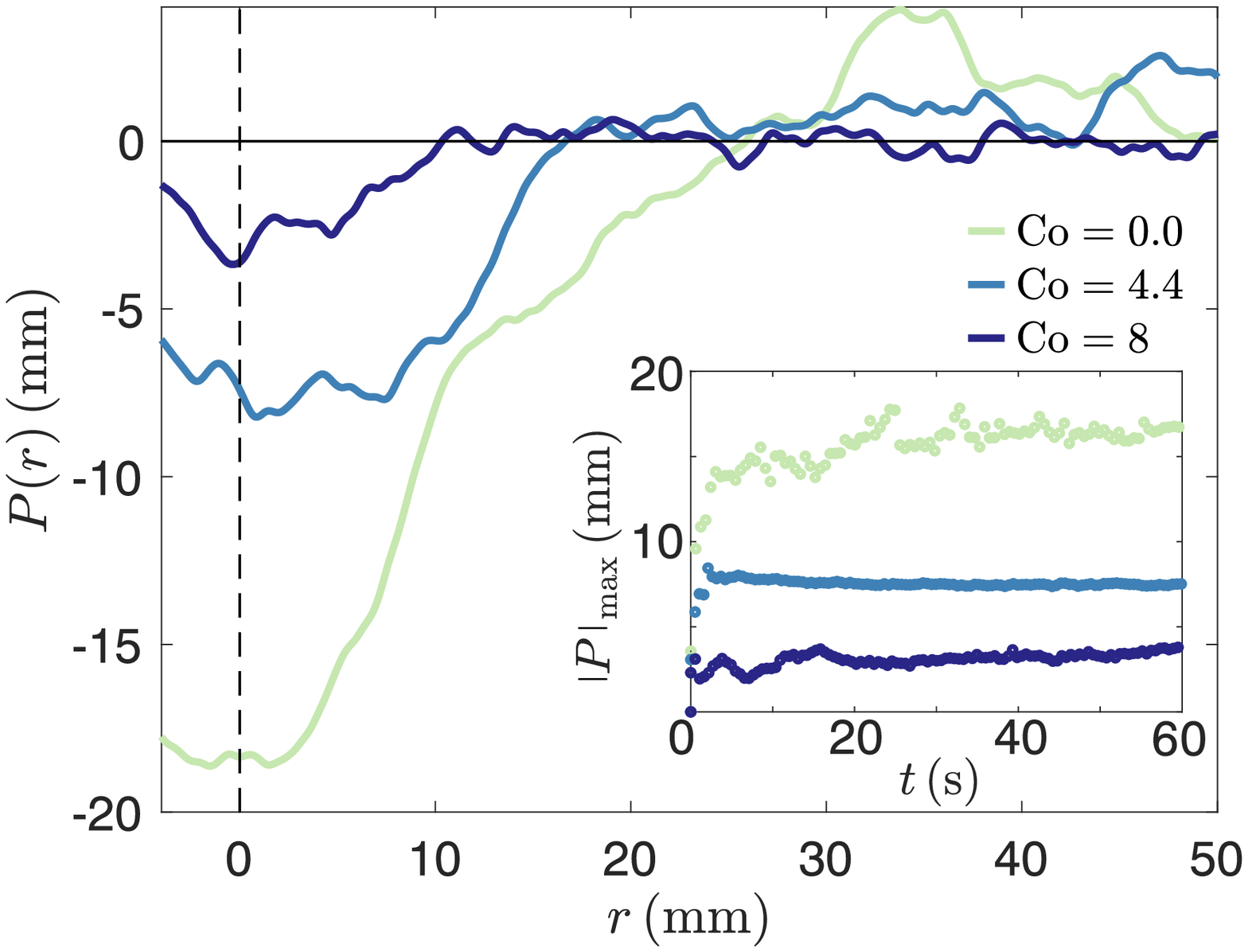}}
\subfigure[]{\includegraphics[width = 0.49\textwidth]{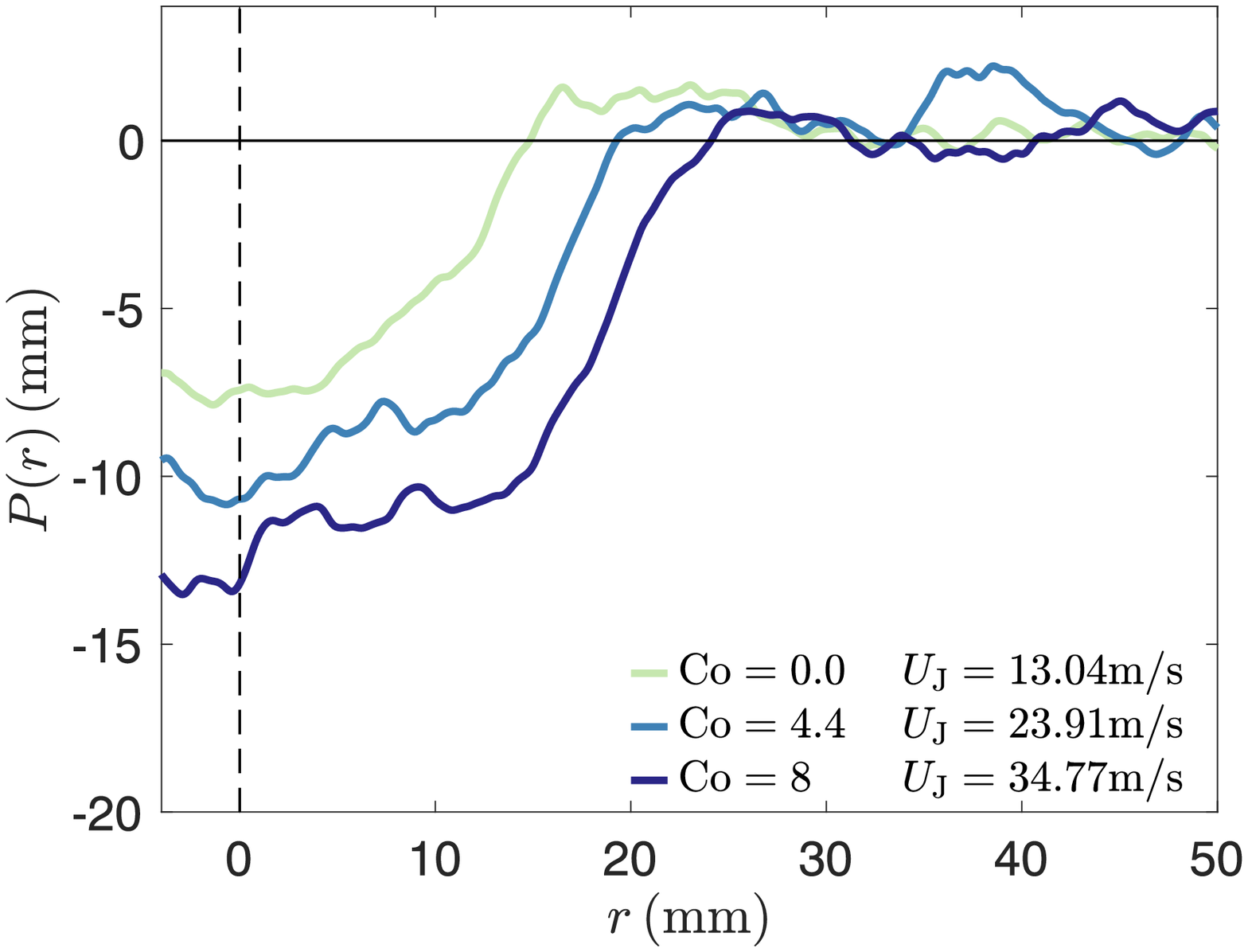}}
\caption{(a) Profiles of craters in dynamical equilibrium for identical jet parameters ($H/D =12.6$; $U_{\rm J}=21.7\,\mathrm{m\,s^{-1}}$) and three different cohesion numbers. Inset: Time-evolution of the depth of the crater at $r=0$. $|P|_{\rm max}=|P(r=0)|$ reaches a constant value after some time showing when the crater has reached a dynamical equilibrium, at which point the profiles reported here were imaged. (b) Profiles of the craters for $\rm{Sh_{\rm \ell, c} - Sh_{\rm \ell, c}^* \approx 0.03}$ and three values of $\mathrm{Co}$. The distance is $H/D =12.6$ and the individual velocities at the nozzle $U_{\rm J}$ are listed alongside the relevant $\mathrm{Co}$ values. In both plots, the solid black horizontal line indicates the granular bed at the start of the experiment, and the dotted vertical line indicates the central axis.}
\label{fig:Figure5_CraterMorphology}
\end{figure}

We report in figure \ref{fig:Figure5_CraterMorphology}(a) the difference observed when varying the inter-particle cohesion, keeping all other parameters constant. For $H/D = 12.6$ and a fixed jet velocity $U_{\rm J} = 21.7\,{\rm m\,s^{-1}}$, the cohesionless case displays a type II crater while moderately cohesive (${\rm Co=4.4}$) and more cohesive (${\rm Co=8}$) granular materials display a type I crater instead. For the most cohesive grains considered here, our entire range of tests ($U_{\rm J}$ up to $35\, {\rm m\,s^{-1}}$ for $H/D=12.6$ and up to $43.5\, {\rm m\,s^{-1}}$ for $H/D=26.7$) showed only type I craters. Besides, we observe that stronger cohesion leads to smaller craters. 

The asymptotic shape of the crater provides information on the final state only. To probe the dynamics of formation of the crater, we report in the inset of figure \ref{fig:Figure5_CraterMorphology}(a) examples of the time-evolution of the crater depth ($|P|_{\rm max}$) at $r=0$, \textit{i.e.}, at the center where the depth is maximum. Note that because of the size of the grains and the dynamics involved, the value at $r=0$ may present some fluctuations or may not be the maximum at a given time. Nevertheless, it provides interesting information on the dynamics and, in particular, how cohesive force can modify the timescale of the formation of the crater. Interestingly, our experiments reveal that the timescale needed to reach the asymptotic crater seems similar, or even reduced, compared to cohesionless grains. This observation can be due to the fact that once the cohesive grains are eroded from the granular bed, they are going to be transported away from the crater much faster. The timescale required to erode the grain itself, \textit{i.e.}, overcoming gravity and cohesion effects does not seem to be modified by the cohesive force.

Instead of considering the jet velocity $U_{\rm J}$, which does not account for the increase of the erosion threshold due to the cohesion, we now consider the cohesive Shields number. The transition between the two regimes (type I and type II) and the evolution of the properties of the crater with non-cohesive grains has been considered by Badr \textit{et al.} \cite{badr2016crater}. They have reported that the distance of the jet Shields number from its critical value for erosion, $\rm{Sh_{\rm J} - Sh_{\rm J}^*}$, provides a framework to discuss the shape of the crater, and is a relevant parameter for the transport rate of eroded grains \cite{andreotti2013granular}. Above a particular value, $\rm{Sh_{\rm J} - Sh_{\rm J}^*} \geq 26$ for their experimental configuration, the shape of the crater evolves from type I to type II. Note that the precise value of the transition between the different craters depends on the size of the grains. Here, we consider larger grains ($d=800\,\mu{\rm m}$) and still observe this transition between the types I and II of craters for our experiments with cohesionless grains. However, for cohesive grains, no transition is observed in this range of parameters since only type I craters are obtained. Using the local cohesive Shields number, we plot in figure \ref{fig:Figure5_CraterMorphology}(b) the morphology of the crater for a fixed deviation from the threshold Shields number, $\rm{Sh_{\rm \ell, c} - Sh_{\rm \ell, c}^* \approx 0.03}$, three different cohesion numbers $\mathrm{Co}$, and $H/D = 12.6$. To account for the change in the threshold Shields number due to the cohesion, the jet velocity has to be increased in this case. The craters observed for these three experimental configurations are all of type I. While figure \ref{fig:Figure5_CraterMorphology}(a) shows experiments performed with the same jet velocity and three different cohesions, the velocity must be changed to obtain similar deviation from the threshold Shields number. The curves appear to have flipped as a consequence of a much larger velocity in the more cohesive cases [figure \ref{fig:Figure5_CraterMorphology}(b)]. The value of the parameter $\rm{Sh_{\rm \ell, c} - Sh_{\rm \ell, c}^*}$ has been shown to be useful in determining the morphology for cohesionless grains \cite{badr2016crater}. Here, even if the distance to the threshold cohesive Shields number $\rm{Sh_{\rm \ell, c} - Sh_{\rm \ell, c}^*}$ is the same, the asymptotic morphology of the craters are not the same in the presence of cohesion. This difference in morphology could be understood through the difference between the erosion threshold and the transport rate. Whereas the onset of erosion is correctly described by accounting for inter-particle cohesion, as developed in the previous section, once a grain has been eroded and is suspended in the air, it is not subject anymore to cohesive force. Different expressions have been obtained for the saturated flux of grains transported by turbulent flows (see \textit{e.g.}, \cite{lajeunesse2010bed,duran2012numerical,andreotti2013granular}), but overall they all show that the transport rate is proportional to $\rm{Sh_{\rm \ell, c} - Sh_{\rm \ell, c}^*}$. For a cohesive granular bed, the erosion threshold will be given by the local cohesive Shields number ${\rm Sh}_{\ell,c}^*$ described in the previous section. However, since the particles eroded are not subject to a cohesive force anymore, the transport rate will be proportional to ${\rm Sh}_{\ell}-{\rm Sh}_{\ell}^*$. Therefore, for a given value of the jet velocity just above the cohesive threshold, the cohesive grains eroded may be transported much further. A detailed description of the transport process of cohesive grains and the role of both the cohesive and cohesionless Shields number is beyond the scope of the present work but will be an interesting aspect to consider in the future, for instance in a unidirectional and homogeneous tangential flow.

A last qualitative feature observed only with cohesive grains occurs once the jet is turned off. In the case of cohesionless grains (${\rm Co}=0$), once the flow stops, if the slope of the crater is larger than the avalanche angle, the grains relax, and the crater collapse on itself, leading to the accumulation of grains at the bottom of the crater. The type II crater relaxes to display only one angle, giving rise to a conical shape. For large cohesion, the crater relaxes much lesser, and the final profile after the cessation of the jet is not very different from the profile of the crater in dynamic equilibrium. Further investigation on the relaxation of the crater is needed.

\section{Conclusion}

\begin{figure}
\centering
\includegraphics[width = \textwidth]{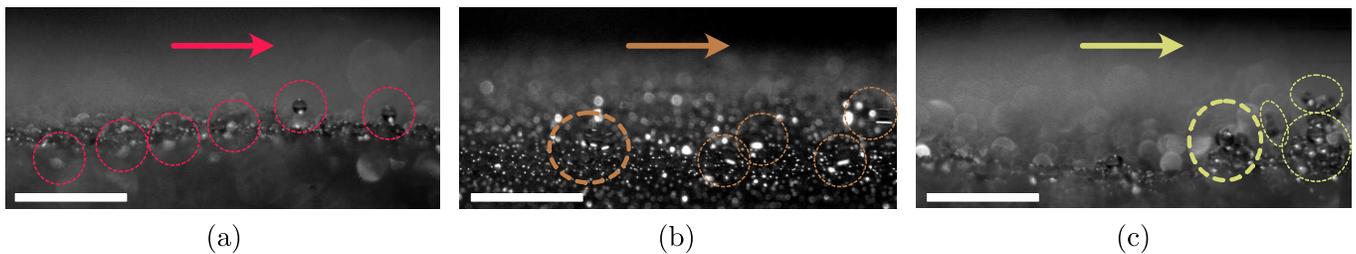}
\caption{Example of visualization of the erosion at the particle scale for (a) a cohesionless granular bed, (b) a cohesive granular bed (${\rm Co}=7.5$) and (c) a larger inter-particle cohesion (${\rm Co}=15$). (a) For the cohesionless case, erosion is seen one grain at a time, and the erosion and transport of one grain are shown captured every 0.05 s. (b) For cohesive grains (${\rm Co}=7.5$), erosion is seen in individual grains as well as small clumps that later separate, here into four individual grains over 0.025 s. (c) For larger cohesion, here ${\rm Co}=15$, we primarily see the erosion of larger clumps breaking into smaller clumps as shown here for two successive images 0.025 s apart. In all figures, the scale bar is 5 mm, and the arrows show the direction of transport. The corresponding movies are available in Supplementary Materials.}
\label{fig:Figure12_LocalView}
\end{figure}

In this article, we have described the role of inter-particle cohesion on the erosion of a granular bed by an impinging turbulent jet. Using model cohesive grains for which the cohesive force can be finely tuned, we have been able to consider the influence of the cohesion number ${\rm Co}$ keeping all the other parameters constant. Our experiments revealed that the inter-particle forces delay the erosion process. Nevertheless, introducing the cohesive force, in addition to the gravity, in the definition of the Shields number allows defining a cohesive Shields number that captured the onset of erosion for the range of cohesion numbers considered here.

Beyond the erosion threshold, the transport process and the asymptotic crater observed also depend on the cohesion. Interestingly, the evolution of the crater shape is not trivial with the cohesion number. A notable feature is that once a grain is eroded, it is not subject to cohesive force anymore. As a result, for a given distance to the threshold Shields number, once the cohesive force has been broken, the grains can be transported much further instead of falling back in the crater, leading to larger craters.

This study is a first step to achieving a fine physical description of the erosion and transport of cohesive particles. In particular, we report in figure \ref{fig:Figure12_LocalView} high-speed imaging of the transport process at the grain scale. Whereas cohesionless grains are transported individually, as shown in figure \ref{fig:Figure12_LocalView}(a), the erosion and transport of cohesive particles are much more complex. In particular, for large enough cohesion, as illustrated in figures \ref{fig:Figure12_LocalView}(b) and \ref{fig:Figure12_LocalView}(c), the erosion could occur through clusters of particles that are later fragmented in individual grains once they leave the viscous sublayer. The value of $\alpha$ obtained in equation (\ref{eq:localShields_coh}) for the local cohesive Shields numbers assumed the erosion grain by grain. This erosion mechanism will likely not be valid for more cohesive grains where the possibility to erode clusters of particles may become larger than the possibility to erode individual particles. In particular, the largest cohesion considered here for the erosion threshold, ${\rm Co}=10$ in figure \ref{fig:Figure4_ShieldsLocal}(a), shows a slight departure from the model developed here. Future works will consider the threshold of erosion of isolated particles versus clusters when varying the inter-particle cohesion.

The results presented in this study are relevant to develop a better understanding of environmental processes involving the erosion and transport of cohesive particles, for instance, due to van der Waals force \cite{ternat2008erosion,fernandes2019investigating} or to biocohesion, which have been shown to modify the formation of bedforms and the onset of erosion \cite{malarkey2015pervasive,chen2017hindered}. The cohesion-controlled granular material used here \cite{gans2020cohesion} is an exciting model material to finely isolate the role of cohesion on erosion and transport processes while keeping other parameters constant.

\appendix

\section{Structure of the turbulent jet} \label{sec:jetStructure}

\begin{figure}
\centering
\includegraphics[width = 0.6\textwidth]{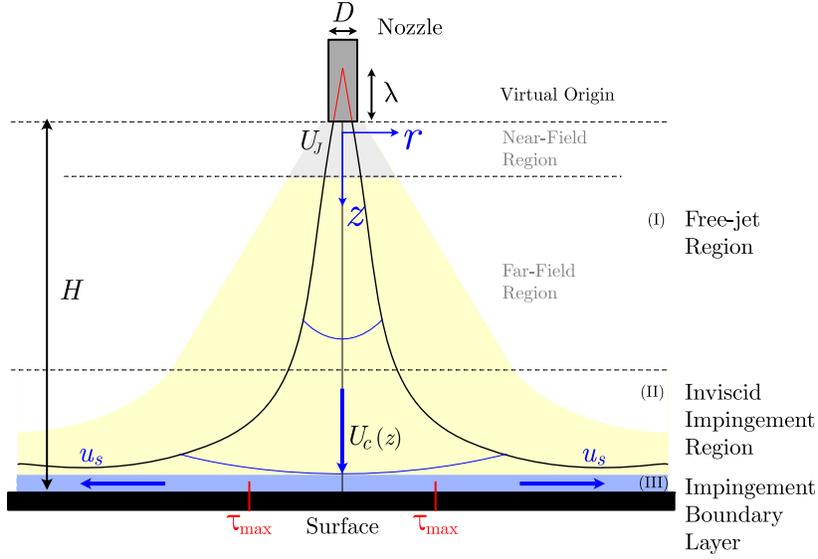}
\caption{Schematic of an axisymmetric turbulent jet impinging on a solid surface showing three main regions: (I) a free-jet region, (II) an inviscid impingement region and (III) an impingement boundary layer \cite{059.Phares-2000}. The average outlet velocity at the nozzle is $U_{\rm J}$. The virtual origin $\lambda$, the centerline velocity $U_{\rm c}(z)$ and the location of maximum shear stress $\tau_{\rm max}$ are also shown. In regions (I) and (II) once the jet becomes self-similar, the centerline velocity scales as $U_{\rm c}(z) \propto U_{\rm J}/z$. We estimate the velocity at the surface as the velocity at the bottom of region (II). For $r \simeq b_{1/2}$, we have $u_s = u_{\rm s, \, max} \approx U_c (z=H) $.} 
\label{fig:Figure2_TurbJet}
\end{figure}

The normal impingement of a turbulent axisymmetric jet on a flat surface is a complex configuration that has been considered in various studies, for instance by Phares \textit{et al.} \cite{059.Phares-2000}. In this appendic, we recall the key results that are used in this study. 

The jet-surface system can be roughly divided into three regions, as shown in the schematic in figure \ref{fig:Figure2_TurbJet}: (I) a free-jet region, (II) an inviscid impingement region and (III) an impingement boundary layer at the surface. In the free jet region (I), we use the same treatment as if there were no impinged surface. In this region, after some initial development the jet has a self-similar structure where the mean velocity is a function of the distance $z$ from the outlet of the nozzle and of the radial distance to the centerline $r$ \cite{davidson2015turbulence,pope2000turbulent}. The region where the jet is still developing, is shown shaded in grey in all figures that are plotted as a function of distance. The inviscid impingement region (II) is where the effects of the wall cause primary deflections to the streamlines, which transition from being mainly normal to being mainly parallel to the solid surface.

In this article, we denote the center-line velocity $U(z,r=0) = U_c(z)$ when the jet is fully developed for regions I and II using the same expression (see figure \ref{fig:Figure2_TurbJet}). The centerline velocity $U_c$ is higher for a larger velocity at the nozzle $U_{\rm J}$, and reduces when the distance from the nozzle $z$ increases, so that $U_c(z) \propto U_{\rm J} / z$. We also introduce a virtual origin, $\lambda$, which corresponds to the distance to the apparent point source for the jet from the nozzle \cite{kotsovinos1976note,badr2014erosion,brunier2017erosion}. Therefore, at a distance $z=H$ from the nozzle, the centerline velocity is given by:
\begin{equation}
U_c (z=H, r=0) = U_{\rm J} \frac{\kappa^{1/2}}{\varepsilon_o} \left(\frac{H + \lambda}{D}\right)^{-1}
\end{equation} where $\kappa$ and $\varepsilon_o$ are constants associated with the turbulent jet. More specifically, $\kappa$ is the jet momentum that measures the strength of the jet, and $\varepsilon_o$ is its virtual kinematic viscosity \cite{059.Phares-2000}. The ratio, $\varepsilon_o / \kappa^{1/2}$ is empirical and has been estimated in various studies, exhibiting significant variation depending on the experimental conditions (for instance between 0.016 and 0.018 in Ref. \cite{schlichting2016boundary}).

The other important parameter in the self-similar region of the jet is a measure of how fast the jet spreads. This is quantified though the half-width  $b_{1/2}$, which corresponds to the radial distance away from the centerline where the velocity is half the centerline velocity. Using the parameters provided by Phares \textit{et al.} \cite{059.Phares-2000} for an axisymmetric turbulent jet, we have $b_{1/2} = 5.27(\varepsilon_o / \kappa^{1/2})z$. Thus, the jet spreads slowly downstream as $b_{1/2} \simeq 0.09\,z$ using $\varepsilon_0/K^{1/2} \simeq 0.017$. 

At the impingement surface, the centerline velocity stream produces a stagnation point. Using a laminar boundary layer model, Phares \textit{et al.} also provided an expression for the magnitude and the radial width of the maximum shear stress caused by the impinging jet on the surface \cite{059.Phares-2000}:
\begin{equation} \label{eq:maxShear}
\tau_{\rm max, \, z=H} = 44.6 \, \rho_a \, {U_{\rm J}}^2 \, {{\rm Re}_{\rm J}}^{-1/2} \left(\frac{H}{D}\right)^{-2}\quad {\rm at} \quad r (\tau_{\rm max}) = 0.09\, H
\end{equation} The analysis with the virtual origin can be considered to only provide a small correction to the dimensionless distance \cite{badr2014erosion}. The locations of the maximum shear is a little distance away from the axis of central impingement as illustrated in figure \ref{fig:Figure2_TurbJet}. Note that in our experiments, the location of the first grains eroded is not at the vertical of the jet but slightly outward, in agreement with the structure of the turbulent jet described above.

A final point used in this study is that the velocity at the bottom of region (II) is approximated as the velocity at the surface, \textit{i.e.}, $U(z=H,r)$= $u_{s}$. If we consider the radial profile, the surface velocity reaches a maximum at $r \simeq b_{1/2}$, where it is comparable to the center-line velocity at $z=H$ so that $u_{\rm s,\, max}(z=H) \simeq U_{c}(z=H)$ \cite{059.Phares-2000}. As we travel far away from the centerline, this surface velocity is expected to decrease, as the jet spreads radially. Since in this study we are primarily interested in the onset of the erosion process, we make use of both $u_{\rm s,\, max}$ and $\tau_{\rm max}$ in our analysis. 

\bibliographystyle{ieeetr}
\bibliography{Biblio_JET_Cohesive}

\end{document}